\documentclass[aip,amsmath,amssymb, reprint]{revtex4-1}
\usepackage{physics}
\usepackage{amsmath}
\usepackage{mathptmx,bm}
\usepackage{graphicx}
\usepackage{dcolumn}
\usepackage{tabularx}
\usepackage{epsf}
\usepackage{color}
\usepackage{hyperref}
\hypersetup{breaklinks,colorlinks,linkcolor=blue,citecolor=blue,urlcolor=blue}

\usepackage{epstopdf}%
\usepackage{verbatim}
\usepackage[english]{babel}
\newcommand{\bea}{\begin{eqnarray}}
	\newcommand{\eea}{\end{eqnarray}}
\usepackage{amsfonts}
\usepackage{graphicx}
\usepackage{dcolumn}
\usepackage{bm}
\usepackage[utf8]{inputenc}
\usepackage[T1]{fontenc}
\usepackage{mathptmx}
\usepackage{etoolbox}

\makeatletter
\def\@email#1#2{%
	\endgroup
	\patchcmd{\titleblock@produce}
	{\frontmatter@RRAPformat}
	{\frontmatter@RRAPformat{\produce@RRAP{*#1\href{mailto:#2}{#2}}}\frontmatter@RRAPformat}
	{}{}
}%
\makeatother

\begin{document}

\title{Simulations of Heat Transport in Single-Molecule Junctions: Investigations of the Thermal Diode Effect}

\author{Jonathan J. Wang}
\affiliation{Chemical Physics Theory Group, Department of Chemistry, University of Toronto, 80 Saint George St., Toronto, Ontario M5S 3H6, Canada}

\author{Jie Gong}
\affiliation{Department of Mechanical Engineering, Carnegie Mellon University, Pittsburgh, PA, 15213, USA}

\author{Alan J. H. McGaughey}
\affiliation{Department of Mechanical Engineering, Carnegie Mellon University, Pittsburgh, PA, 15213, USA}

\author{Dvira Segal}
\affiliation{Chemical Physics Theory Group, Department of Chemistry, University of Toronto, 80 Saint George St., Toronto, Ontario M5S 3H6, Canada}

\affiliation{Department of Physics, University of Toronto, 60 Saint George St., Toronto, Ontario M5S 1A7, Canada}
\email{dvira.segal@utoronto.ca}

\date{\today}

\begin{abstract}
With the objective to understand microscopic principles governing thermal energy flow in nanojunctions, we study phononic heat transport through metal-molecule-metal junctions 
using classical molecular dynamics (MD) simulations.
Considering a single-molecule gold-alkanedithiol-gold junction, we first focus on aspects of method development and compare two techniques for calculating thermal conductance: (i) The Reverse Nonequilibrium MD (RNEMD) method, where heat is inputted and extracted at a constant rate from opposite metals. In this case, the thermal conductance is calculated from the nonequilibrium temperature profile that is created on the junction. (ii) The Approach-to-Equilibrium MD (AEMD) method, with the thermal conductance of the junction obtained from the equilibration dynamics of the metals. 
%
In both methods, simulations of alkane chains of growing size display an approximate length-independence of the thermal conductance, with calculated values matching computational and experimental studies.
The RNEMD and AEMD computational methods  offer different insights on thermal transport, and we discuss their relative benefits and shortcomings.
Assessing the potential application of molecular junctions as thermal diodes, the alkane junctions are made spatially asymmetric by modifying their contact regions with the bulk, either by using distinct endgroups or by replacing one of the Au contacts by Ag. Anharmonicity is built into the system within the molecular force-field.
Using the RNEMD method, we show that, while the temperature profile strongly varies (compared to the gold-alkanedithiol-gold junctions) due to these structural modifications, the thermal diode effect is inconsequential in these systems---unless one goes to very large thermal biases. 
This finding suggests that one should seek molecules with considerable internal anharmonic effects for developing nonlinear thermal devices. 

%
\end{abstract}

\maketitle

\section{Introduction}

The function, performance, and stability of electronic, plasmonic, thermal, and thermoelectric devices fundamentally rely on their heat conduction properties and energy dissipation pathways \cite{Pop10,Baowen12,Luo13,Rev14,Leitner15,RevA,Yoon20,RevG,BaowenR21}.
%
Unlike electrons, which are directly controlled by electrostatic and electromagnetic fields, phonons, the quanta of vibrational energy, are not feasibly manipulated by external driving forces. 
Can we control and direct vibrational energy flow at the nanoscale, down to the level of a single molecule? Before addressing this question, we must first understand the relation between molecular structure and the ensuing thermal transport properties.

Recent studies of phononic heat conduction in {\it single-molecule} junctions were focused on quasi one-dimensional (1D) organic molecules. Specifically, 
since alkane chains are poor conductors of charge carriers, one can safely assume that their thermal conductance is dominated by their nuclear motion and neglect the contribution of electrons. 
Phononic thermal transport in alkane chains of 2-10 repeating units (and sometimes longer) was simulated in Refs. \citenum{Dvira2003,Pawel11,Pauly16,Pauly18,Roya19,Nitzan20, Lu2021,Nitzan22} with different techniques (classical or quantum)
and varying degree of details (one-dimensional or three-dimensional models, explicit metals or Langevin baths), with measurements reported in Refs. \citenum{CuiExp19,GotsmannExp19}.
Experimental studies were also performed on self-assembled monolayers (SAMs) of alkane molecules \cite{Wang06,Dlott07,Cahill12,GotsmannExp14,Shub15,Shub17} with atomistic simulations reported in Refs. \citenum{Hu10,Luo10,Kikugawa14,Diamond17,Shub15,Shub17}. 
Both single-molecule and SAM junctions were generally shown to support ballistic (non-dissipative) heat transport that was 
approximately length-independent in long enough chains (typically beyond 10 units).
Given their simple structure and the ensuing ballistic thermal behavior, alkane chains serve as a testbed for developing computational methodologies for single-molecule phonon heat transport.



Towards the long-term objective of deciphering
the structure-function question in phonon heat flow at the nanoscale, in this study we focus on two objectives: 
(i) method development and benchmarking of computational techniques, (ii) modelling and simulations of nonlinear thermal devices, specifically single-molecule thermal diodes. 
Considering alkane-based junctions, we focus on the atomistic contribution to thermal energy transport, also referred to as phononic heat transport
(in the bulk, which is the source of thermal energy in our work, the phonon description is valid).
Compared to quantum calculations \cite{Quantum1,Quantum2}, classical MD simulations can feasibly include anharmonic interactions, which could be influential in molecules at high temperatures.
Moreover, for Au-alkane-Au nanojunctions it was found that quantum statistics played a small role around room temperatures \cite{Lu2021}, justifying simulations based on classical MD.

We perform atomistic-classical nonequilibrium MD simulations of Au-alkanedithiol-Au junctions and related-modified systems using LAMMPS \cite{LAMMPS}, 
and study their thermal conduction as a function of length, temperature bias, and contact properties.
The first part of the paper, Sec. \ref{sec:Methods} is devoted to aspects of method development and benchmarking. 
Here, our objective is to find what (possibly complementary) microscopic information different computational techniques convey, 
their benefits and shortcomings. We utilize two methodologies for extracting the phonon conductance of molecular junctions:
In the first method, which we discuss in Sec. \ref{sec:RNEMD}, we implement reverse nonequilibrium molecular dynamics (RNEMD) simulations. 
Here, kinetic  energy is added to one Au lead, and removed from the other side, both at a constant rate.
In response, a temperature gradient develops across the junction, allowing us to calculate the thermal conductance of the junction.
This method was implemented to describe heat conduction in a silicon-polyethylene-silicon single-molecule junction\cite{Pawel11} and in Au-SAM-Au systems \cite{Luo10,Shub15,Shub17}. 
The second method, discussed in Sec. \ref{sec:AEMD} is referred to as the approach-to-equilibrium molecular dynamics (AEMD) method. This tool was recently applied to investigate the thermal resistance of  molecular junctions based on ab-initio quantum dynamics \cite{AEMD}. 
In this approach, the thermal conductance is determined from the equilibration dynamics of the bulk metals by monitoring their changing temperatures. 
In Sec. \ref{sec:disc}, thermal conductances 
obtained from these two computational approaches are compared to other studies and to experiments and discussed.

Continuing to applications, the main nontrivial question that we probe in the subsequent Sec. \ref{sec:Diode} is whether an alkane-based junction could materialize the thermal diode effect by making it spatially asymmetric through modifications to its boundaries.
Thermal diodes are devices that conduct heat asymmetrically upon reversal of the temperature bias \cite{WalkerRev11,DiodeRev17,Terraneo02,Baowen04,Bambi06}. Based on minimal models, see e.g. Refs. \citenum{SB1,SB2}, it was argued that a diode effect can be generated at the nanoscale when two conditions are met: 
The junction is spatially-asymmetric and anharmonic effects are at play.
We incorporate asymmetry in two distinct ways: (i) We modify one of the endgroups to enhance asymmetry, as well as anharmonicity.
(ii) We build junctions with mismatched metal contacts.
In both cases, anharmonicity is included in the force-field. 
Our analysis reveals that while the temperature profile generated in the molecule is sensitive to spatial asymmetry, the ensuing thermal diode effect is insignificant in alkane-based junctions unless the temperature bias is made very large ($T_h-T_c> 200$ K, with $T_{h,c}$ the temperatures of the opposite metals).
Sec. \ref{sec:mism} brings predictions on the linear conductance of homogeneous and mismatched junctions. 
We summarize our work in Sec. \ref{sec:Summary}, further suggesting directions for future work. 

\begin{figure*}[tb]
\centering
\includegraphics[width=2\columnwidth]{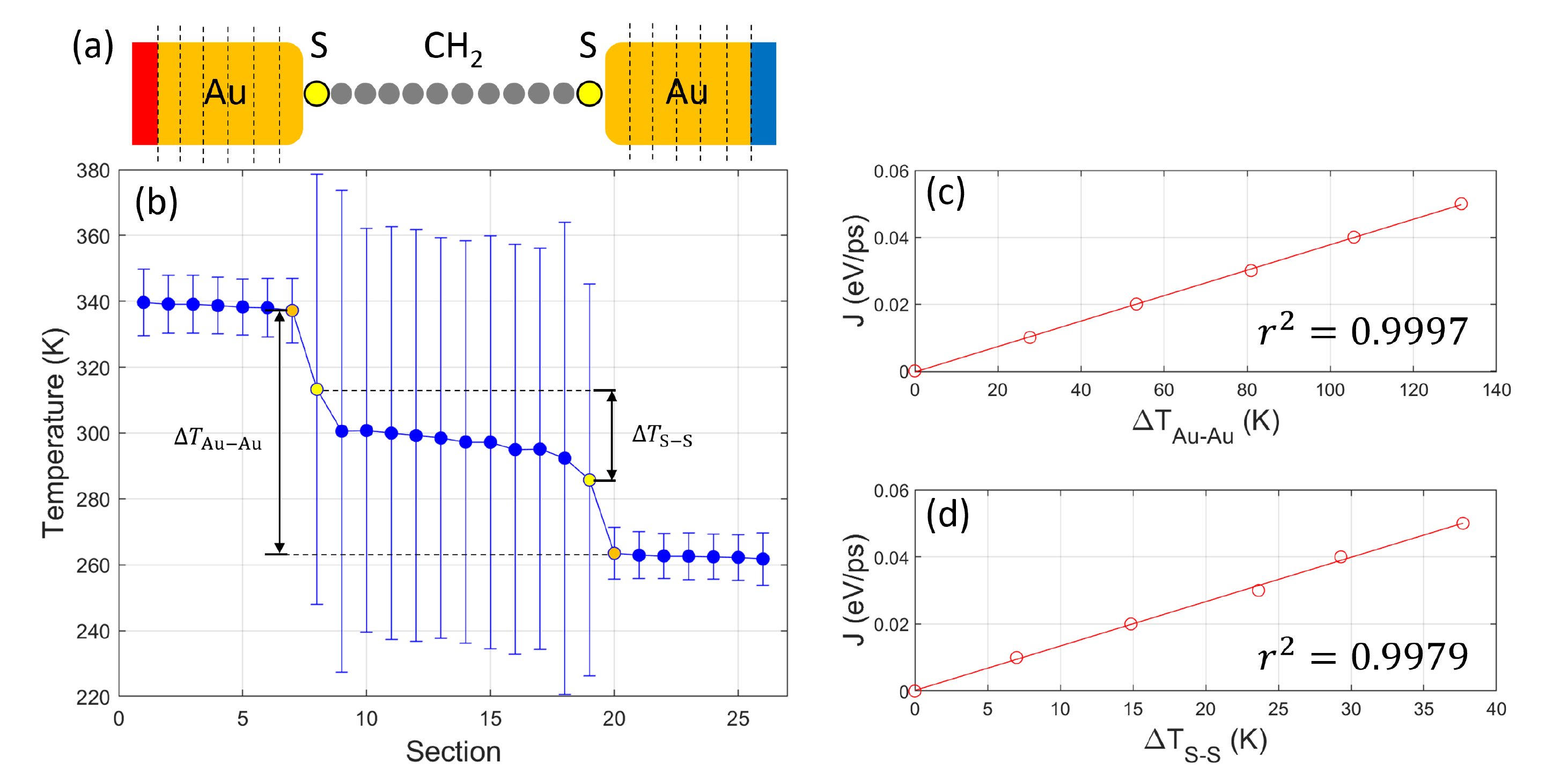} 
\caption{
Simulation results from the RNEMD method.
\textbf{(a)} A graphical representation of a single-molecule junction; regions in the metals over which the temperature is evaluated are separated by dashed lines.
\textbf{(b)} The temperature profile obtained in RNEMD simulations, shown for $N_{\text{C}}=10$ under the heat current $J=0.02$ eV/ps. Each data point represents a time-averaged temperature of Au sections and individual atoms in the chain as displayed in (a). We define two temperature biases, $\Delta T_{\text{Au-Au}}$ and $\Delta T_{\text{S-S}}$, evaluated as the difference between either gold or yellow data points, respectively.
\textbf{(c), (d)} Imposed heat current against the resulting temperature bias (c) $\Delta T_{\text{Au-Au}}$ and (d) $\Delta T_{\text{S-S}}$ with molecular length $N_{\text{C}}=10$.}
\label{NEMDmethod}
\end{figure*}

\begin{figure}[hbt]
    \centering
    \includegraphics[width=0.9\columnwidth]{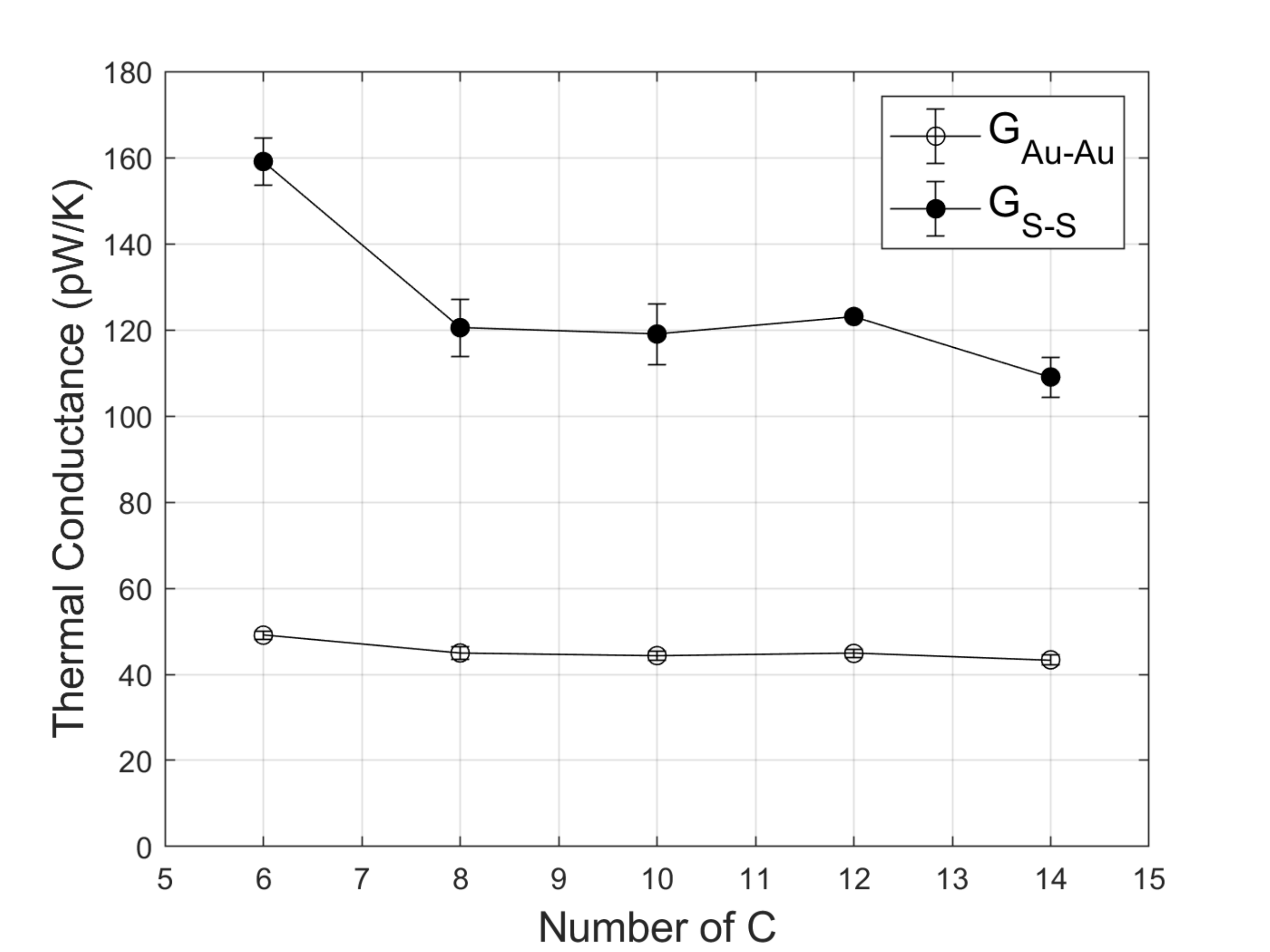} 
\caption{RNEMD thermal conductance as function of molecular length displaying $G_{\text{Au-Au}}$ and  $G_{\text{S-S}}$.
Data was obtained by performing simulations as presented in Fig. \ref{NEMDmethod} while varying the molecular length: 
For each junction, the heat current is set at $J=0.02$ eV/ps, and we extract the temperature differences $\Delta T_{\text {Au-Au}}$ and $\Delta T_{\text {S-S}}$ to evaluate the respective conductance.
Each data point was averaged over three independent simulations.
}
\label{NEMDlength}
\end{figure}

\section{Simulation techniques}
\label{sec:Methods}

We discuss here the two simulation methods that we implemented and tested for studying thermal conductance with LAMMPS: The reverse  nonequilibrium MD, which is based on steady-state simulations, and the approach-to-equilibrium MD, which relies on transient dynamics.


\subsection{Reverse Nonequilibrium Molecular Dynamics}
\label{sec:RNEMD}


The RNEMD method employs a reverse procedure for determining 
the current-thermal bias (or specifically the thermal conductance), compared to the direct nonequilibrium MD method \cite{Nitzan20,Nitzan22,Lu2021,Hu10,Luo10,Kikugawa14,Diamond17,NEMD19}. In RNEMD, one equilibrates the total system to a certain temperature (300 K in this work),  then imposes the heat current as a parameter. In turn, the temperature profile in the junction is obtained in the nonequilibrium steady-state limit. The heat current is given by the rate of adding (extracting) kinetic energy into (from) atoms in the hot (cold) baths. By splitting the system into sections, a temperature profile resultant of the energy flow is obtained by calculating the kinetic energy on sections of the metals and the molecule. 

We define the temperature bias {\it on the junction} as $\Delta T_{\text{M-M'}}$, with M$=$Au, Ag identifying the metal contact; the temperatures are evaluated on the metal, but close to the molecule.
We also characterize an {\it intrinsic molecular} temperature difference, which is defined between the two endgroups. In
an alkanedithiol junction, this temperature bias is measured S-to-S, $\Delta T_{\text{S-S}}$. We illustrate these two temperature differences in Fig. \ref{NEMDmethod}.

These biases are used for the determination of two thermal conductance values,  $G_{\text{M-M'}}$ and $G_{\text{S-S}}$, using the linear response expression,
\bea
G = \frac{J}{\Delta T}.
\eea
$G_{\text{M-M'}}$ corresponds to experimental measurements of thermal conductance in molecular junctions \cite{CuiExp19,GotsmannExp19}, since there, temperatures are practically measured at the metal contacts. 
In contrast, while $G_{\text{S-S}}$ is affected by the molecule being hybridized to the metal contacts, it provides deeper insights into the thermal transport behavior within the molecule itself. 

We present in Fig.~\ref{NEMDmethod} a graphical representation of the junction and its partition into sections (a), the resultant nonequilibrium temperature profile under a certain value of the heat current $J$ (b),
and a demonstration of the approximate linear relationship between the heat current and the temperature difference (c)-(d), where again one needs to remember that the heat current is fixed in simulations and the temperature difference is the calculated-simulated value.
Appendix A provides details of the setup and the simulation procedure. The force field and its parameters are described in Appendix C.

Focusing on Fig.~\ref{NEMDmethod}(b), which exemplifies
the characteristics of the temperature profile, we point out that its form agrees with the literature \cite{Shub15,Nitzan20}. 
Note the large temperature fluctuations on the S and C atoms, compared to the Au sections. This is expected since each gold section consists of many (270) atoms, allowing better averaging and reduced fluctuations within the leads---compared to molecular sections, which are made of single atoms. 
Additional observations are: (i) A small gradient is presented within the Au sections. 
(ii) The temperature profile is approximately spatially-symmetric, reflecting the structural symmetry.
(iii) The lion's share of the temperature bias drops at the interfaces between the gold leads and the sulfur atoms. Only a small temperature gradient develops on the alkane chain itself. This behavior matches the fact that the harmonic part of the molecular force field dominates over anharmonic contributions, thus transport on the molecule is close to ballistic \cite{Lebo67} (anharmonic interactions are included in the molecular dihedral and Au-interface interactions). Metal-molecule interfacial thermal resistance \cite{BaowenR} is thus the main source of resistance in Au-alkanedithiol-Au junctions, as we also show in Fig.~\ref{NEMDlength}.

The results in Fig.~\ref{NEMDmethod}(c)-(d)
demonstrate a linear relationship between the heat current and the temperature difference, allowing us to extract both $G_{\text{Au-Au}}$
and $G_{\text{S-S}}$ as linear response coefficients.
The molecule-length dependency of both $G_{\text{Au-Au}}$ and $G_{\text{S-S}}$ are shown by Fig.~\ref{NEMDlength}. While $G$ appears to be higher at short lengths,
the thermal conductance is approximately constant for chains with $N_\text{C}=8-12$ units, yet manifesting a small decline for longer chains.
Based on the dominance of the harmonic part in the alkane force-field, alkane-based junctions are expected to follow the ballistic transport behavior, with minimal inelastic-dissipative effects. The ballistic mechanism in our system is reflected by the thermal conductance being almost independent with length, and it is supported by experiment \cite{CuiExp19}. However, other studies of alkane chains discovered that finite-size effects can cause the thermal conductance to peak at short molecular lengths \cite{Dvira2003,GotsmannExp14}.  

Our calculated Au-to-Au thermal conductance, $G_{\text{Au-Au}}$, is in a good agreement with the literature, both in terms of values and trends, see Sec. \ref{sec:disc}. In what follows, unless otherwise stated, we use $G_{\text{M-M'}}$ as the relevant measure for the junction's conductance, with the temperature difference evaluated from the metal atoms at the boundaries. 

Examining in Fig.~\ref{NEMDlength} the intrinsic conductance, $G_{\text{S-S}}$, we observe similar trends to $G_{\text{Au-Au}}$, with the conductance close to saturating for long chains.
As expected, the intrinsic molecular conductance is greater than the junction's value,  $G_{\text{S-S}}>G_{\text{Au-Au}}$,
reflecting the contribution of contact resistance to the latter. It is remarkable to note the extent of the suppression of conductance due to the interface resistance. This reinforces the argument that transport is close to being ballistic (nonresistive) in the molecule \cite{RevA}. 


\subsection{Approach-to-Equilibrium Molecular Dynamics}
\label{sec:AEMD}

The AEMD method relies on the phenomenological Newton's law of cooling. In this method, the thermal conductance is determined from the rate of thermal equilibration of the metal leads. Starting with a nonequilibrium condition, the two separate metal leads are prepared at distinct temperatures, hot and cold. Once the metals are attached via the molecule, energy flows between the metals through the molecule, approaching a global equilibrium state. By monitoring the bulk temperature while relaxing to equilibrium, one can determine the thermal conductance of the junction.

\begin{figure*}[htbp]
\centering
\includegraphics[width=2\columnwidth]{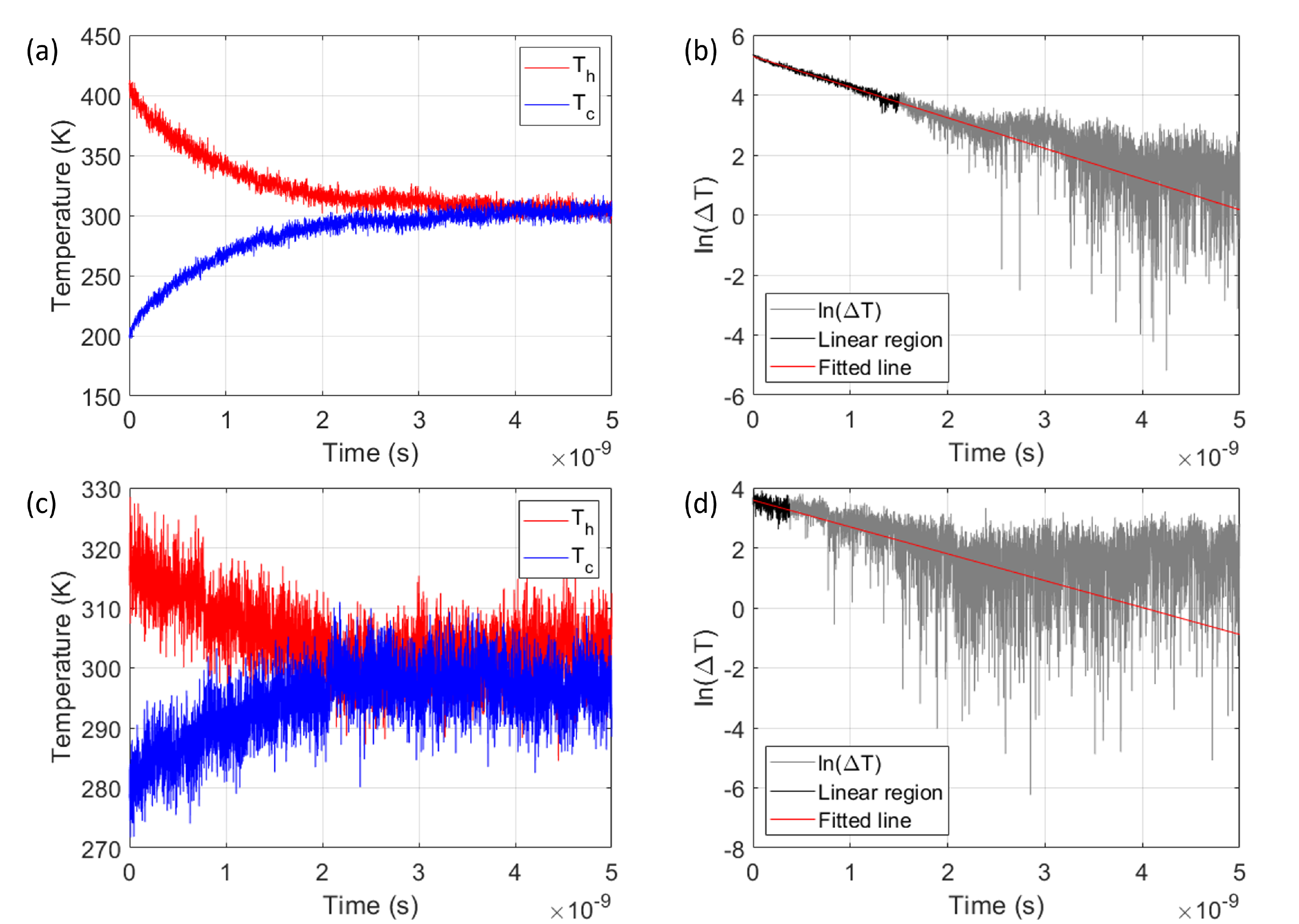} 
\caption{
AEMD simulations of thermal conductance.
Temperatures of the hot ($T_h$) and cold ($T_c$) gold regions over AEMD simulation time, and plots of $\ln(\Delta T)$ over time for \textbf{(a), (b)} initial temperature bias of 200 K and \textbf{(c), (d)} 50 K. The decay timescale $\tau$ used in the determination of the thermal conductance, Eq. (\ref{eq:GAE}), is obtained from a linear fit of the $\ln(\Delta T)$ plot before equilibration.
Panels {\bf (c)}, {\bf (d)} highlight the limitation of the AEMD method at low temperature biases, as the exponential decay towards equilibrium is not sufficiently manifested.
Parameters are $N_{\text{C}}=10$ in both cases and $\bar T(0)=300$ K.}
\label{AEMDmethod}
\end{figure*}
%
\begin{figure}[hbtp]
\centering
\includegraphics[width=\columnwidth]{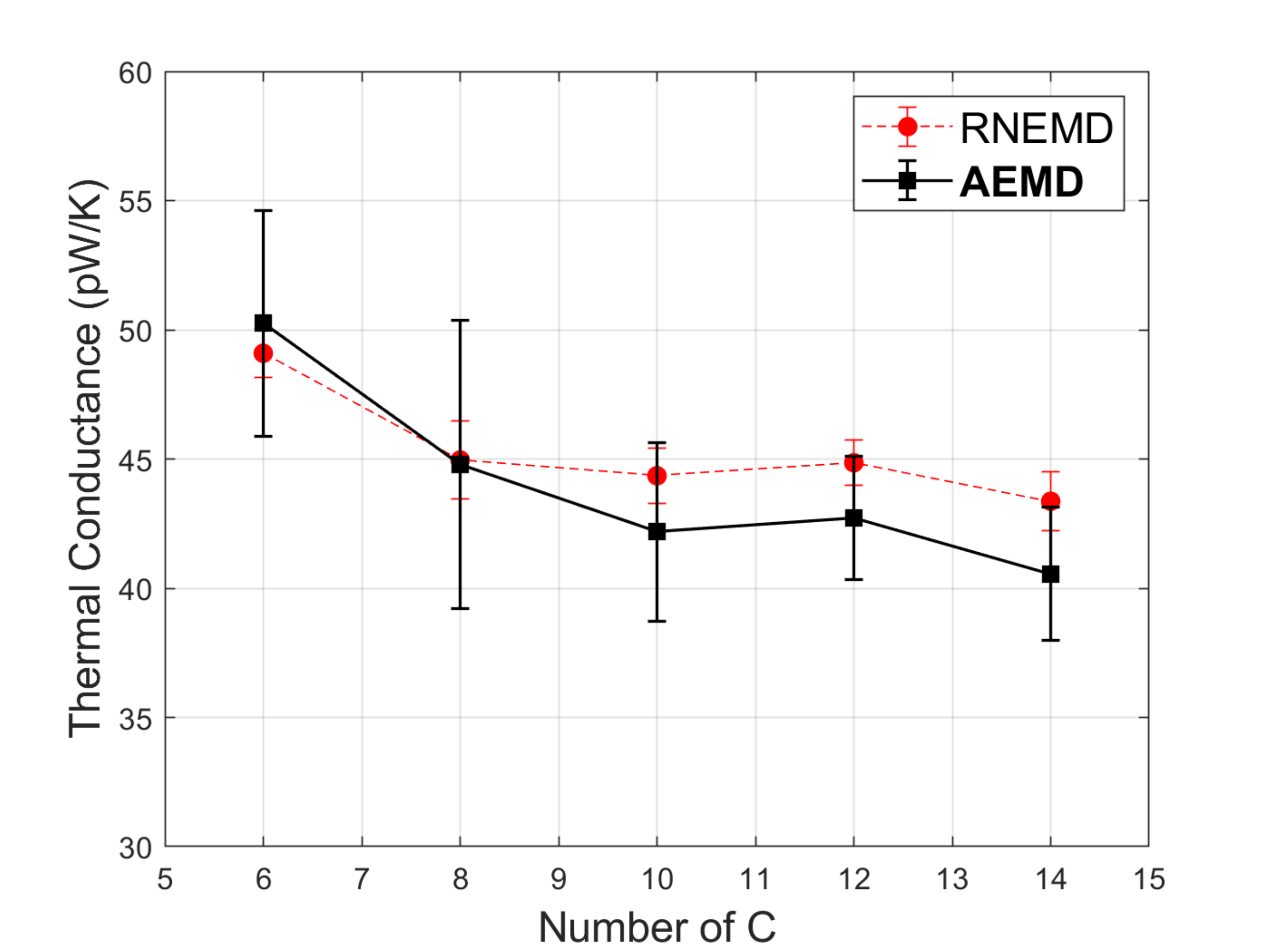} 
\caption{AEMD simulations of thermal conductance as a function of molecular length ($\square$). RNEMD results from Fig. \ref{NEMDlength}(a)  are overlayed for comparison ($\circ$). Parameters are $T_h(0)=400$ K, $T_c(0)=200$ K.
Each AEMD data point was obtained by averaging over 5 runs.}
\label{AEMDlength}
\end{figure}
%
Newton's cooling law for each metal lead is given by
\bea
C\frac{dT_h(t)}{dt} &=& -G\left[T_h(t)-T_c(t)\right ]
\nonumber\\
C\frac{dT_c(t)}{dt} &=& -G\left[T_c(t)-T_h(t)\right],
\label{eq:New}
\eea
where $C$ is the heat capacity of the lead. 
Using the Dulong–Petit law, which is justified in the classical MD simulations, the heat capacity is approximated as $C=3k_BN_{\text{Au}}$, 
where $N$ is the number of moving gold atoms in the lead (see Appendices A and B) atoms and $k_B$ the Boltzmann constant.
$G$ is the thermal conductance of the junction. Defining the temperature difference as $\Delta T=T_h-T_c$, the two equations (\ref{eq:New}) can be combined,
\bea
\frac{d\Delta T(t)}{dt} = -\frac{2G}{C} \Delta T(t).
\eea
The solution to this differential equation is given by
\bea 
\Delta T(t) = \Delta T(0)\; e^{-t/\tau},
\label{eq:DTsol}
\eea
where $\tau\equiv C/2G$ is the equilibration time. For later use, we also define the averaged temperature as $\bar T(t)= [T_h(t)+ T_c(t)]/2$, which is set at $\bar T(0)=300$ K for the AEMD simulations.

We display examples of AEMD simulations in Fig.~\ref{AEMDmethod}
with initial temperature biases of 200 K (top) and 50 K (bottom).
Raw data with the changing temperatures of the metals 
is shown in Fig.~\ref{AEMDmethod}(a),(c). In Fig.~\ref{AEMDmethod}(b),(d), we present
$\ln (\Delta T)$ as a function of time in accord with Eq. (\ref{eq:DTsol}). While at short times an exponential decay is observed (linear decay in the log scale), at long time the system approaches equilibrium and it no longer follows Newton's cooling law.
Focusing on the appropriate exponential-decay regime, colored in black in Fig.~\ref{AEMDmethod}(b),(d), one can extract the slope $1/\tau$ and evaluate the thermal conductance from the AEMD method as
\bea
G = \frac{C}{2\tau}.
\label{eq:GAE}
\eea
%
In Fig.~\ref{AEMDlength} we show that thermal conductances from the AEMD method are in agreement with RNEMD results for $G_{\text{Au-Au}}$.
While larger finite-size effects seem to affect the conductance in AEMD, results are within repeated simulation error and trends observed in the two methods are similar.
The AEMD method thus similarly exposes an approximate ballistic transport behavior, compounded with finite-size effects \cite{Dvira2003}. 

While Fig.~\ref{AEMDlength} suggests that RNEMD and the AEMD are both valid, there are several limiting aspects of the AEMD method that make it less favorable than RNEMD. These aspects are discussed in the next section, concluding that the RNEMD method is a more robust technique. 
Thus, our method of choice in applications, Sec. \ref{sec:Diode}, is RNEMD.


\subsection{Comparison: methods and experiments}
\label{sec:disc}

The results from the RNEMD and AEMD methods match well, as we show in Fig. \ref{AEMDlength}. While both methods display what seems to be finite-size effects of the conductance in short chains, ballistic transport is more evident in RNEMD results for longer systems.

We now discuss the pros and cons of the two methods.
Beginning with the RNEMD method, the determination of thermal conductance in this technique is based on measuring the temperature profile along the chain, in steady state. This method however is not limited to linear-response, as one can more generally interrogate with RNEMD the relationship between $\Delta T$ and the heat current, as we do in the next section. On the down side, production runs with RNEMD are relatively long, at $\approx 25$ ns, an order of magnitude longer than AEMD simulations (see Fig. \ref{AEMDmethod}(a),(c)).
Indeed, the main appealing aspect of the AEMD is its shorter simulation time with decreased computational cost.

Despite its short simulation time, the AEMD method suffers from several deficiencies.
Newton's cooling law is phenomenological and it relies on several assumptions including that $G$ is temperature-independent and that the decay behavior is controlled by a single timescale.
%
Regarding the former, this issue will be of a concern in systems for which $G$ varies with temperature with problems expected to be manifested when going to high temperature biases. 
A second limitation of AEMD concerns studies at small biases. As noted in Fig.~\ref{AEMDlength}, simulations of $G$  over length were obtained using a temperature bias of 200 K, which  is  larger than what experiments nowadays can feasibly allow \cite{CuiExp19,GotsmannExp19}. Thus, we attempted to determine $G$ with AEMD at a lower bias, comparable to the aforementioned experiments. Raw data is displayed in Fig.~\ref{AEMDmethod}(c)-(d). We immediately note the extent of fluctuations in the temperature trajectory due to the small temperature bias in Fig.~\ref{AEMDmethod}(c), compared to Fig.~\ref{AEMDmethod}(a). These fluctuations, and the short equilibration time limit the adoption of the AEMD cooling equations. Fig.~\ref{AEMDmethod}(d) shows that the linear region is short and noisy, thus extracting the timescale $\tau$ is imprecise. Indeed, our fitting procedure lead to a $G$ value of 35 pW/K, which is outside the range of results in Fig.~\ref{AEMDlength}. 
Additional constraints on the AEMD method are that it assumes that there is no spatial variation of the temperature in the bulk, and that it is not proper to be used on long molecular systems because the phenomenological cooling expression assumes a single exponential decay. In contrast, in long chains ($N=14$ according to our tests)  oscillations decorate the exponential decay, indicating more involved dynamics than a single exponential decay.
 
Based on both RNEMD and AEMD methods, the thermal conductance of single-molecule S-alkane-S junction of $N_C\approx 10$ sites is at 40-45 pW/K. 
In Table~\ref{Table1}, we compile relevant thermal conductance values from the literature for alkane junctions, both computational with different methods, as well as experimental; our obtained results (row \#1) should be compared to relevant experimental studies (rows \#7-\#8), as well as to computations on the same system (rows \#2-\#5). SAMs bring comparable values when consideration is given to the per area aspect of their thermal conductance.
	
\begin{table*}[htbp]
	\begin{center}
	\caption{Table of thermal conductance values from theoretical and experimental studies}
	\footnotesize
	\begin{tabular}{|m{2em}|m{7em}|m{5em}|m{5em}|m{8em}|m{14em}|m{8em}|}
		\hline
		\# & \bf    Reported conductance  & \bf Molecule type & \bf Interface type & \bf Single-molecule/SAM & \bf Experiment/Theory & \bf Authors \\ 
		\hline
		1. &\bf  40-45 pW/K & S-alkane-S & Au & Single-molecule & Theory (Classical MD) & \bf This work\\
		\hline
2. &		20 pW/K & S-alkane-S & Au & Single-molecule & Theory (Classical MD) & Sharony \textit{et al}.\cite{Nitzan20}\\
		\hline
		3. & 39 pW/K & S-alkane-S & Au & Single-molecule & Theory (Classical MD) & Majumdar \textit{et al}.\cite{Shub17}  \\
		\hline
		4. & 30-35 pW/K & S-alkane-S & Au & Single-molecule & Theory (Semiclassical MD) & Li \textit{et al}.\cite{Lu2021}\\
		\hline
		5. & 35-45 pW/K & S-alkane-S & Au & Single-molecule & Theory (Quantum NEGF) & Kl\"ockner \textit{et al.}\cite{Pauly16}\\
		\hline
		6. & 180 pW/K & Alkane & Si & Single-molecule & Theory (Classical MD) & Sasikumar \textit{et al}.\cite{Pawel11}  \\
		\hline
		7. & 25 pW/K & S-alkane-S & Au & Single-molecule & Experiment & Cui \textit{et al}.\cite{CuiExp19} \\
		\hline
		8. & 40 pW/K & S-alkane-S & Au & Single-molecule & Experiment & Mosso \textit{et al}.\cite{GotsmannExp19} \\
		\hline
		\hline
		9. & 400 MW/m$^2$K & S-alkane-S & Au & SAM & Theory (Classical MD) & Luo \textit{et al}.\cite{Luo10}\\
		\hline
		10. & 260 MW/m$^2$K & Alkane & Si & SAM & Theory (First-principles AEMD) & Duong \textit{et al}.\cite{AEMD} \\
		\hline
		11. & 18 MW/m$^2$K & Alkane & Diamond & SAM & Theory (Classical MD) & Wang \textit{et al}.\cite{Diamond17}\\
		\hline
		12. & 60 MW/m$^2$K & S-alkane-S & Au & SAM & Experiment & Majumdar \textit{et al}.\cite{Shub15} \\
		\hline
		13. & 16 pW/K & S-alkane & Au, Si & SAM & Experiment & Meier \textit{et al}.\cite{GotsmannExp14}\\
		\hline
		
	\end{tabular}
\label{Table1}
\end{center}
\footnotesize *Reported $G$ are those  most relevant to our system, such as the approximate thermal conductance of a 
10-carbon chain, if the authors examined length dependence.
\end{table*}

\begin{figure}[hbt!]
\centering
\includegraphics[width=\columnwidth]{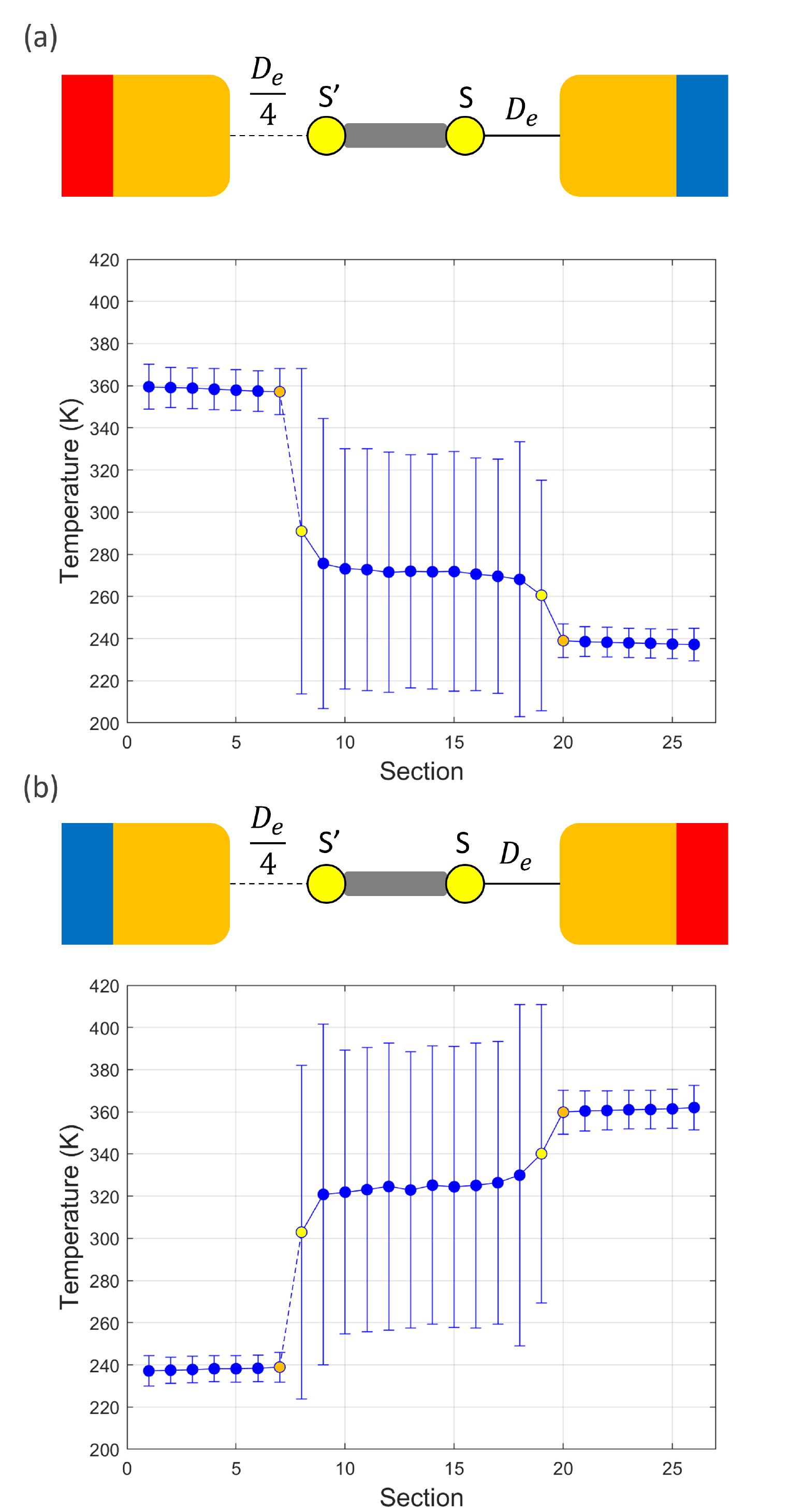} 
\caption{Asymmetric junctions created by manipulating the dissociation energy of endgroups connected to Au. 
 \textbf{(a)} and \textbf{(b)} show two identical, asymmetric junctions with imposed heat current flowing in opposite directions. 
 The  respective temperature profiles from RNEMD are displayed, manifesting a large jump in temperature at the weaker contact.
Parameters are heat current of $J=0.02$ eV/ps and $N_{\text{C}}=10$.}
\label{Diode_setup}
\end{figure}
%
\begin{figure*}[hbt]
	\centering
\includegraphics[width=2\columnwidth]{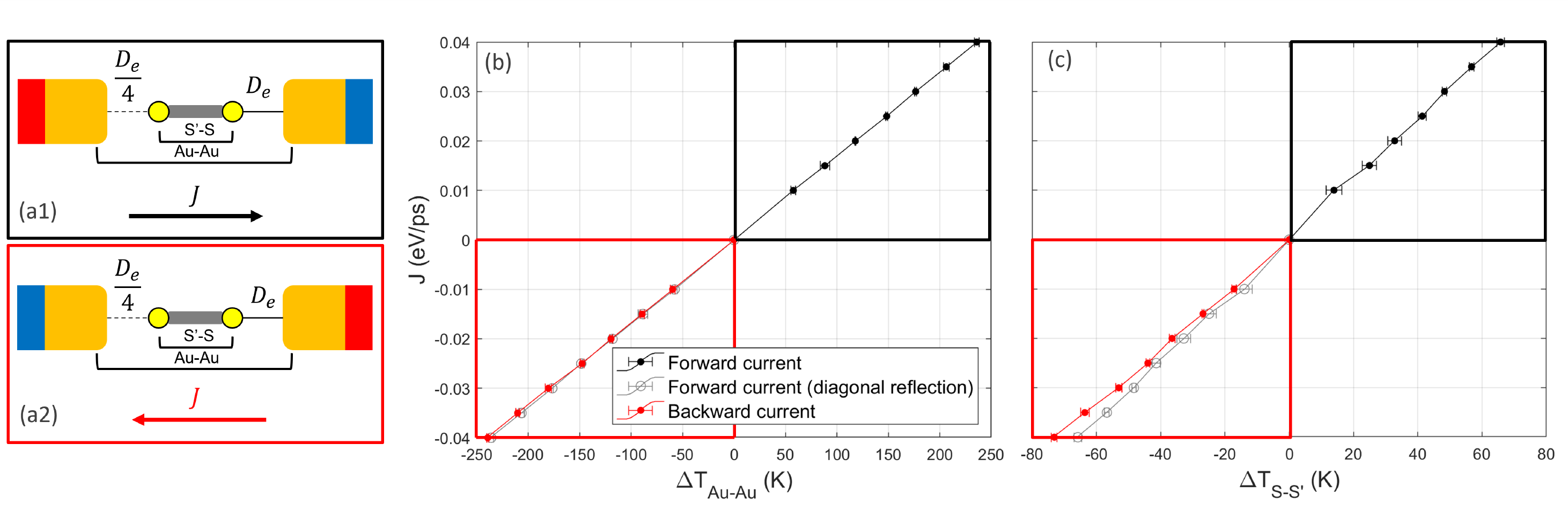} 
\caption{Analysis of the thermal diode effect in junctions with endgroup asymmetry.
{\bf (a1)}-{\bf (a2) } Graphical representation of molecular junctions in the (a1) forward and (a2) backward current direction.  
{\bf (b)}-{\bf (c)  }Imposed forward (black) and backward (red) currents against (b) $\Delta T_{\text{Au-Au}}$ and {\bf (c)} $\Delta T_{\text{S-S}}$. 
The forward current is also diagonally projected on the left-bottom quadrature (gray, empty $\circ$). 
In simulations, $N_{\text{C}}=10$, the heat current was varied, and the temperature bias was extracted as explained in Fig. \ref{Diode_setup}.}
	\label{Diode_JDT}
\end{figure*}
\section{Thermal Diode Effect}
\label{sec:Diode}

The development of molecule-based thermal devices is a central objective in nanoscale heat transport, with a particular interest in the realization of thermal diodes (rectifiers) \cite{Baowen12,RevG}. In analogy to electrical diodes, thermal diodes present differences in the magnitude of the heat current when the direction of the applied temperature bias is reversed. 
Early studies identified anharmonicity and spatial asymmetry as two key factors in realizing thermal diodes, and efforts were placed on identifying promising molecular and nanoscale setups \cite{Terraneo02,Baowen04,SB1, SB2, Bambi06}.
More recently, advancements in studies of phononic thermal diodes largely occurred through MD simulations of carbon-based materials \cite{BLD2,BLD3,BLD4,BLD5,BLD6,BLD7}.
In parallel to the quest for realizing molecular-based heat diodes, radiative thermal diodes were explored in hybrid normal-superconducting junctions \cite{rad1,rad2,rad3} and by utilizing metal-insulator phase transitions \cite{Biehs13,Xu15,Fiorino18}; these references are examples of a large literature. Additionally, pump-probe time-dependent studies on asymmetric molecules discovered unidirectional vibrational energy flow, a diode-like effect \cite{Dlott13a, Dlott13b,Tucker19}.
Overall, compared to the electrical analogue, thermal control is hindered by immature technology; studies of thermal diodes may lead to the development of desired nonlinear thermal components \cite{WalkerRev11,DiodeRev17, Yoon20,RevG}.

Our objective in this section is to test whether alkane junctions, which were already studied as thermal conductors, can be modified and used as backbone materials in phononic thermal diodes.
Recalling that anharmonicity and spatial asymmetry are necessary conditions in the thermal diode effect, we note that anharmonicity is built-in into the junction in several places: within dihedral interactions, the Morse potential at the boundary in the Au-S bond, and the contribution of the Lennard-Jones and embedded atom model (EAM) potentials. 
As for structural asymmetry, we  introduce it here either by modifying one of the endgroups, or by replacing one of the metal leads. In both setups, the
thermal transport behavior is examined with the RNEMD method, which allows the investigation of current-temperature bias characteristics far from equilibrium.
%

\subsection{Asymmetric endgroups and the thermal diode effect}
\label{sec:Rend}

Our first proposed setup for single-molecule thermal diodes is depicted in Fig.~\ref{Diode_setup}. 
It includes an alkane chain with endgroups of distinct bond dissociation energies. Specifically, we assume that one endgroup involves a ``normal" Au-S bond, while the potential energy of the opposite endgroup denoted by S' is weakened using a smaller bond dissociation energy, $D_e/4$. We refer to the Au-S and Au-S' contacts as ``strong" and ``weak", respectively. Specifically, the $D_e$ values in the Morse potential are 0.38 eV (strong) and 0.095 eV (weak). As such, we created an asymmetric junction as well as enhanced the role of anharmonicity.
The weaker endgroup could represent an Au-methyl or Au-carboxilic contact.
Since our goal here is to explore different setups towards thermal diode applications, we rely here on this phenomenological and flexible mean for enforcing asymmetry, rather than analyze a concrete endgroup.

For comparison, we first study a symmetric junction where both contacts are weak, and find (simulation results not shown) that the thermal conductances $G_{\text{Au-Au}}$ and $G_{\text{S-S}}$ are in the range of 20-22 pW/K and 96-105 pW/K, respectively. Thus, with weak bonds the junction's conductance $G_{\text{Au-Au}}$ is approximately half of the value obtained with normal S-bond interactions (see Fig. \ref{NEMDlength}). In contrast, and in accord with our expectations, the {\it molecular} conductance takes similar values for the strongly- and weakly-bonded junctions,
$G_{\text{S-S}} \approx G_{\text{S'-S'}} $. 

We exemplify in Fig.~\ref{Diode_setup} the temperature profile generated in a current-carrying nonequilibrium situation under forward and reversed currents of the same magnitude. A large temperature drop occurs on the weakly-coupled contact S'-Au, while a smaller temperature difference falls on the stronger S-Au contact. This is expected; the stronger bond better facilitates thermal transport with a reduced contact resistance.
The temperature on the alkane chain thus lies closer to the temperature of the Au lead to which it strongly couples. 

In Fig.~\ref{Diode_setup}, we enforced identical currents in opposite directions. However, to test the existence and extent of the diode effect one needs to compare $J(\Delta T)$ to $J(-\Delta T)$. This is a nontrivial task under the RNEMD method since in this approach we impose currents and gain the corresponding $\Delta T$ as the dependent variable.

Fig.~\ref{Diode_JDT}(a) illustrates the diode setup. Results are presented in Fig.~\ref{Diode_JDT}(b)-(c) showing the heat currents for the asymmetric $D_e$ junctions as a function of the junction's thermal bias, $\Delta T_{\text{Au-Au}}$, and as a function of the intrinsic bias, $\Delta T_{\text {S-S'}}$. 
Recall that the latter bias corresponds to the temperature difference falling on the molecule, between the two endgroups.

When the system operates as a diode, the current-bias profiles of forward and backward junctions should differ, 
$|J(\Delta T)|\neq |J(-\Delta T)|$. To guide the eye, we project the forward current onto the left-bottom quadrature.  Fig.~\ref{Diode_JDT}(b) displays this plot against $\Delta T_{\text{Au-Au}}$, and we find that the diode effect {\it on the junction} is marginal. 
Thus, though (as we show next) a small diode effect exists on the molecule, transport asymmetry is negligible once taking into account the contribution of contact resistance. 
Given the approximate linear current-bias trend, we can also extract the conductance, which is approximately 27 pW/K.
In Fig.~\ref{Diode_JDT}(c) we present the current against the {\it internal} temperature drop, $\Delta T_{\text{S-S'}}$. In this case,  we do observe a small diode effect: The comparison between current-bias profiles indicate that the current is larger  when the weak bond is coupled to the hot metal, than when the weak bond is connected to the cold bath in accord with Refs. \citenum{SB1,SB2}.
Nevertheless, (i) a diode effect is missing in the {\it junction's} definition. 
(ii) Even when utilizing the intrinsic definition, the diode effect is small. We thus conclude that a diode effect is marginal in asymmetric-endgroup alkane junctions under experimentally-relevant applied biases.


\subsection{Mismatched metals and the diode effect}
\label{sec:Rmetal}

We explore here the development of the diode effect in asymmetric junctions created by replacing one of the gold contacts with a different metal, silver. Silver has a higher Debye temperature than gold, $T_D^{\text{Au}}= 180$ K, while $T_D^{\text{Ag}}= 220$ K.
In Au-SAM-Ag junctions, the mismatch in phononic spectral densities of the different metals was shown experimentally to impact (reduce) thermal transport compared to a Au-SAM-Au junction \cite{Shub15}. It is interesting
to probe this effect in single-molecule junctions, as well as to test whether it could lead to a diode effect under large temperature biases.

In simulations, we replaced one of the gold contacts by silver, changing the atomic mass and the metal-metal interaction (EAM potential). However, we assumed that Au and Ag have identical Morse interaction potentials with the molecular endgroup (S). This assumption allows us to concentrate on the role of the mismatch in the phonon spectra of the metals on the diode effect, rather than compounding it with the impact of different endgroups' interactions.

\begin{figure}[hbt!]
	\centering
\includegraphics[width=\columnwidth]{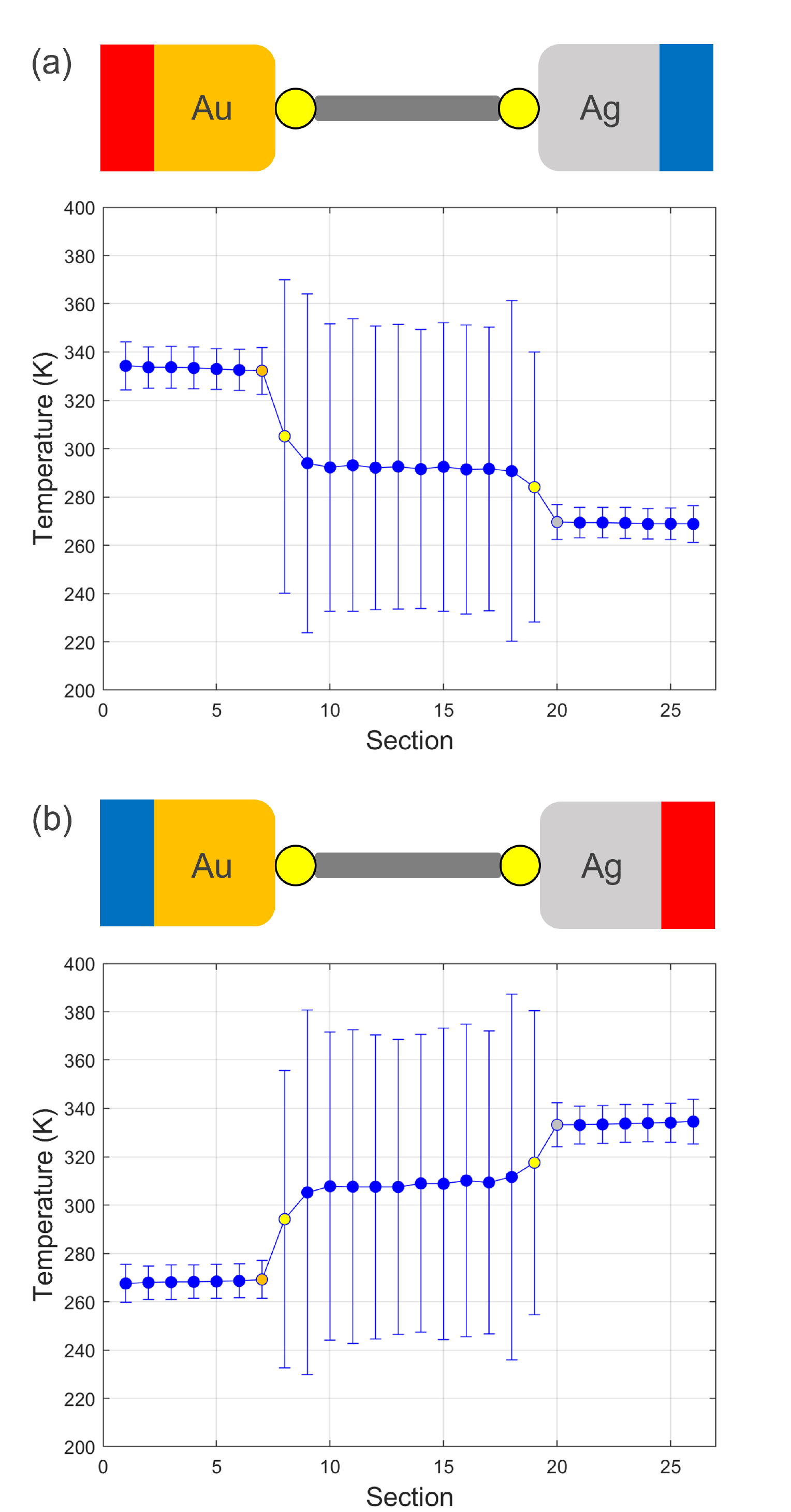} 
\caption{Thermal transport in Au-alkanedithiol-Ag junctions.  \textbf{(a)} and \textbf{(b)} show two identical junctions with heat currents flowing in opposite directions, along with their respective temperature profiles (RNEMD simulations).  Parameters are heat current $J=0.02$ eV/ps and $N_{\text{C}}=10$.}
	\label{Diode_alloy}
\end{figure}

We display in Fig.~\ref{Diode_alloy} a graphical representation of a junction with mismatched metals, along with the  temperature profiles under forward and backward current-carrying conditions. 
Similarly to Fig.~\ref{Diode_setup}, we observe that the averaged temperatures of the backbone C atoms deviate from $T\approx300$ K (the temperature achieved in symmetric junctions, see Fig. \ref{NEMDmethod}).  Instead, the temperature of the molecule lies closer to the temperature of Ag. Comparison with Sec.~\ref{sec:Rend} suggests that Ag, with its higher Debye frequency, allows better phonon transmission to the molecule, thus a smaller contact resistance.
As for the internal temperature profile, we find that it is symmetric under the reversal of the nonequilibrium condition. 

\begin{figure}[hbt!]
    \centering
    \includegraphics[width=\columnwidth]{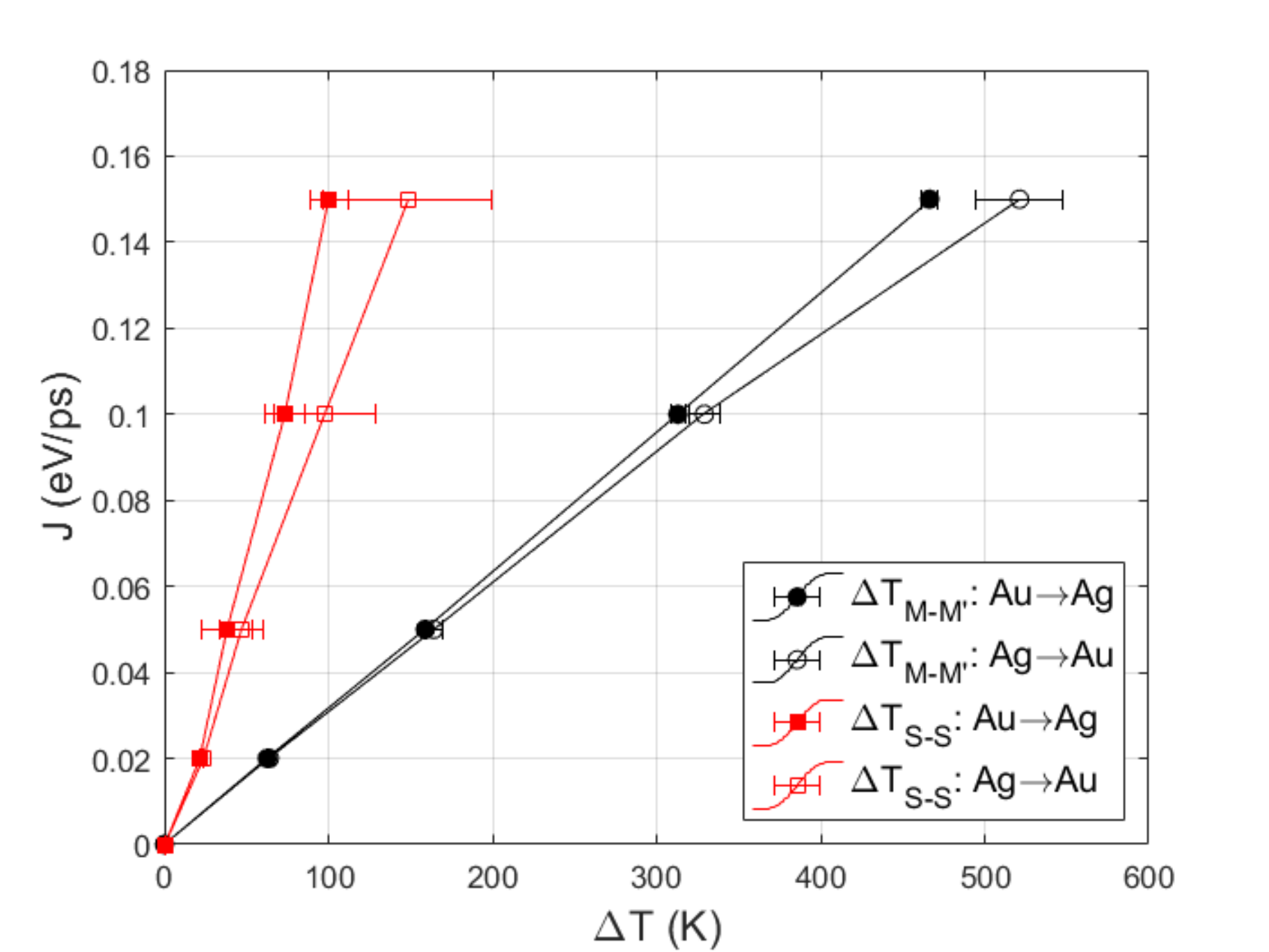} 
    \caption{Study of a diode effect in Au-alkanedithiol-Ag junctions, extending to large heat current and temperature differences. 
    We present the current from the hot Au to the colder Ag (full) and the reversed configuration (empty) as a function of the temperature bias between the metals (circle) or the endgroups (square). 
    The temperature differences $\Delta T_{\text{S-S}}$ and $\Delta T_{\text{M-M'}}$ were calculated based on profiles as in Fig. \ref{Diode_alloy}.
    Simulations were performed with the RNEMD method on an $N_\text{C}=10$ chain.}
    \label{LargeJD}
\end{figure}

We now probe the diode effect in junctions with mismatched metals and compare to the case with asymmetric endgroups. 
The results are presented in Fig. \ref{LargeJD}. 
For small currents, we  confirm a linear trend with $J(+\Delta T)=-J(-\Delta T)$. When we test the behavior under high heat currents, $\Delta T_{\rm M-M'}$ begins to differ when evaluated in the forward or backward directions, with significant deviations once
 $\Delta T_{\text{M-M'}}>200$ K. A clear diode effect shows once the metals are maintained at large temperature differences of 300 K and higher. As for the internal temperature difference, $\Delta T_{\rm S-S}$, we also observe a substantial diode effect forming internally, and at lower temperature biases than when measured at the contacts,  although simulation noise clouds this observation.
These results are in line with the behavior of the asymmetric-endgroup junctions as presented in Fig. \ref{Diode_JDT}, where the internal molecular definition manifested a more noticeable diode effect compared to the marginal metal-to-metal result.

Overall, according to Fig. \ref{LargeJD}, mismatched junctions do not support a substantial diode effect as long as $\Delta T_{\text{M-M'}}<200$ K. While the mismatched metals can support a diode effect, the temperature difference required for its manifestation is outside of what is feasible in experiments, which typically stays in the range of $\Delta T=50-100$ K.


Asymmetry was introduced here in the metals' phonon spectra. We recall a related setup, which does not involve metal contacts:
Motivated by pump-probe experiments in solution \cite{Dlott13a,Dlott13b}, in Ref. \citenum{LeitnerR} a diode-like effect was analyzed in a molecules consisting two distinct endgroup moieties, anthracene and azulene, bridged by a polyethylene glycol oligomer. 
There, a significant diode effect was observed on the molecule due to  the vibrational mismatch between the two endgroups, combined with nonlinear-anharmonic coupling effects.

\subsection{Discussion}
\label{sec:diodeD}

The junctions analyzed in Sec. \ref{sec:Rend}-\ref{sec:Rmetal} were spatially-asymmetric, and they involved anharmonic effects. 
However, only at challenging conditions of high thermal biases could a diode effect be realized. The absence of the diode effect under moderate conditions could be analyzed as follows  \cite{Segal09}:

In both setups (junctions with distinct endgroups or mismatched metals)
and under moderate currents of $J=0.02$ eV/ps
(which is not far from what was used in experiments such as in \cite{CuiExp19}, at about 0.0075 eV/ps), the temperature profile approximately obeys the following relation,
\bea
T_n+\tilde T_n = T_{\text M}+T_{\text M'} = 2\bar T.
\label{eq:Tprofile}
\eea
Here, $T_n$ is the temperature of the $n$th section (part of the metal, or of the molecule) under left-to-right net current, while $\tilde T_n$ is the temperature profile under reversed conditions.
$T_{\text {M,M'}}$ are the temperatures of the metals measured at their boundaries, with  $\bar T$ defined as their average.

We now assume that the heat current can be written as a linear function in the local temperatures,
\bea
J_{\text{M}\to \text{M'}} &=& \sum_n \alpha_n T_n,
\label{eq:Tlinear}
\eea
with $\alpha_n$ as the expansion coefficients.
Based on the fact that $J=0$ at equilibrium, we find that $\sum_n \alpha_n =0$.
Now, to test the diode effect, we turn the bias such that now heat flows from M' towards M. We thus use the temperature profile of the reversed case,
\bea
J_{\text{M'}\to \text{M}} &=& \sum_n \alpha_n \tilde T_n,
\nonumber\\
&=& \sum_n \alpha_n (T_{\text M}+T_{\text M'}-T_n) 
\nonumber\\
&=&-\sum_n \alpha_n T_n,
\eea
which proves the absence of a diode effect. Note that Eq. (\ref{eq:Tlinear}) extends beyond a Landauer-harmonic description and it could account for anharmonic effects and high thermal biases situations, see e.g., Ref. \citenum{Segal09}. 
Thus, as long as (i) the temperature profile obeys the symmetry relation Eq. (\ref{eq:Tprofile}) and the current is linear in the local temperatures, Eq. (\ref{eq:Tlinear}), a diode effect cannot be materialized in the junction.
Figs. \ref{Diode_setup} and \ref{Diode_JDT} for asymmetric endgroups, and Figs. \ref{Diode_alloy} with \ref{LargeJD} for mismatched metals, support this analysis. 



\section{Mismatched junctions: Single-molecule vs self-assembled monolayers}
\label{sec:mism}

\begin{figure}[hbt!]
\centering
\includegraphics[width=\columnwidth]{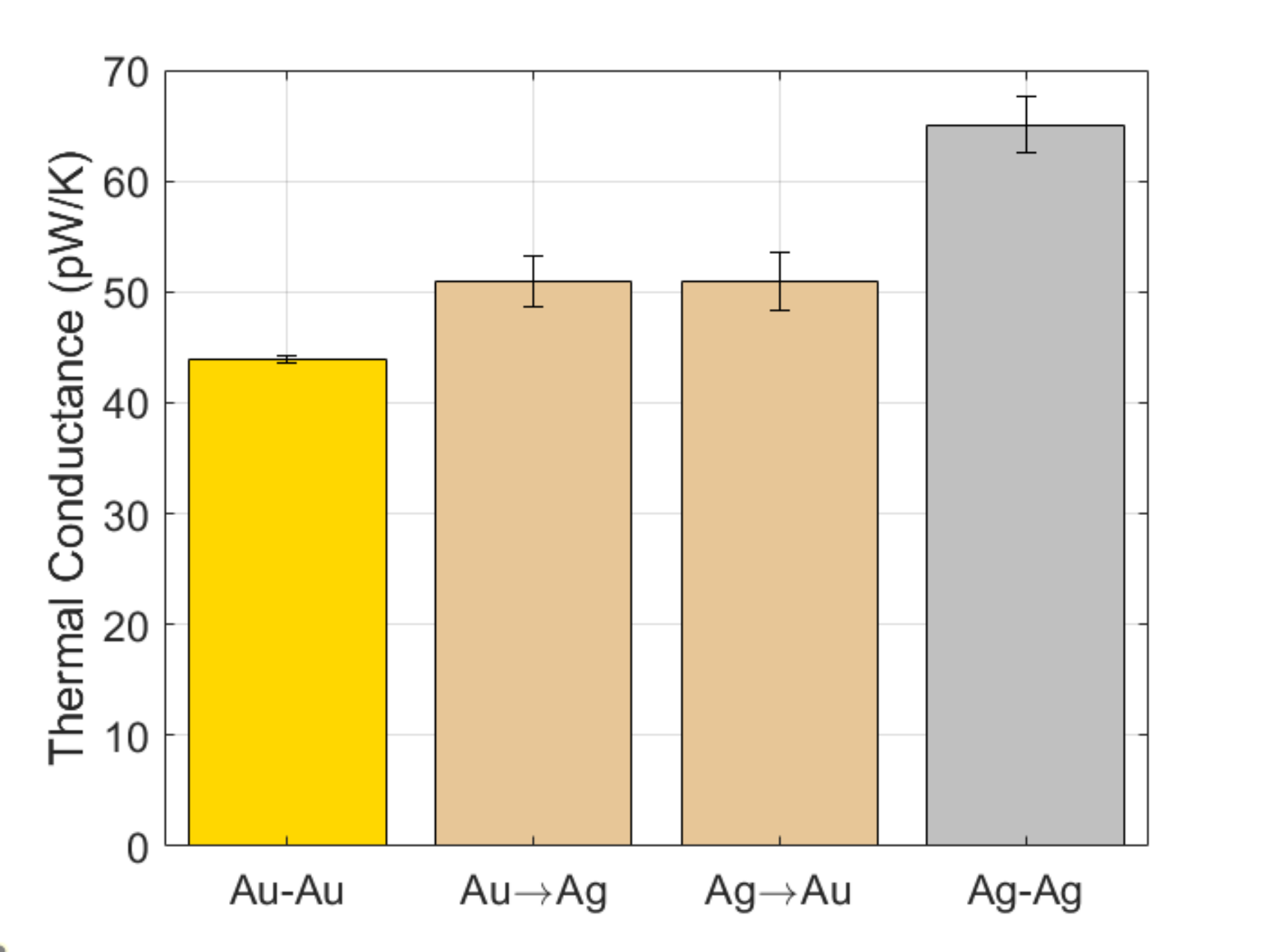} 
\caption{Thermal conductance in M-alkanedithiol-M' junctions with Au and Ag metals.
The arrow in the labeling of asymmetric junctions indicate the direction of net heat flow, from the hot to the cold metal. Parameters are heat current $J=0.02$ eV/ps and $N_{\text{C}}=10$.
}
\label{AuAg_bar}
\end{figure}

We now go back to the linear-response regime and address a question that was previously interrogated for SAMs in Ref. \citenum{Shub15}: What is the role of mismatched metals on the junction's linear response thermal conductance?
We return to the conditions of Fig. \ref{Diode_alloy} where the current was relatively low, yielding $\Delta T_{\text{ M-M'}}\approx 60$ K. In this regime, the heat current vs $\Delta T_{\text {M-M'}}$ trend is about linear, thus one can calculate the linear conductance on the junction from their ratio.
We present in Fig. \ref{AuAg_bar} the thermal conductance of this system, testing junctions with the same metals (Au-alkane-Au and Ag-Alkane-Ag), and with mismatched interfaces (Au-alkane-Ag and Ag-alkane-Au). 
We reveal that the conductance with Ag metals is {\it higher} than with Au.  As for the mismatched  junctions, they support conductances in between those of the homogeneous-metal junctions. This observation, which is based on classical MD simulations, agrees with a harmonic theory that we detail in Appendix D. 

A previous RNEMD study on SAMs compared the conductance of Au-alkanedithiol-Au to  Au-alkanedithiol-Ag \cite{Shub15}, observing a higher thermal conductance in the latter system. Experiments on SAMs \cite{Shub15}, however, showed the opposite trend, with mismatched junctions having conductance {\it below} the homogeneous-gold case. This disagreement was suggested to stem from classical MD simulations activating high-frequency vibrational modes, which in actuality should not participate in heat transport. 
Dealing here with single-molecule junctions rather than SAMs, we expect anharmonic effects in the molecule to be less prominent than in SAMs. In fact, simple harmonic theory (Appendix D) supports our MD predictions with the ordering 
$G_{\text {Ag-Ag}}>G_{\text{Ag-Au}}>G_{\text{Au-Au}}$.
Experimentally verifying this hierarchy for single-molecule alkane chains would provide  a strong validation for the classical MD computational method, as well as an interesting deviation of single-molecule thermal transport from the behavior of SAMs \cite{Shub15}.


\section{Summary}
\label{sec:Summary}

Using classical molecular dynamics simulations, we studied phonon heat transport through single molecule junctions. 
The work addressed two aspects: Method development in the area of classical molecular dynamics, and applications, interrogating whether alkane-based molecular junctions could be modified to serve as thermal diodes.

In the method-development part of this study we employed two classical MD methods. RNEMD is a steady-state approach where one sets the current and obtains the temperature profile across the junction, allowing the calculation of current-bias characteristics and the thermal conductance.
In contrast, in the AEMD method one follows the equilibration dynamics from a nonequilibrium initial condition, receiving the junction's conductance, albeit under some approximations. 

The two methods produced similar results for the thermal conductance of 6 to 14-unit Au-alkanedithiol-Au junctions. Results were in accord with reported experiments, pointing as well to the expected ballistic transport.
Though the AEMD method offers a computational advantage over the RNEMD technique, we concluded that assumptions underlying its analysis restrict its application (e.g., to large thermal biases $\Delta T$) 
deeming the RNEMD method to be a more suitable, flexible and robust method to study thermal transport in molecular junctions.


By constructing spatially-asymmetric junctions, we probed the diode effect in alkane-based junctions far from equilibrium. We constructed two setups: molecules with asymmetric endgroups of strong and weak bonds, and molecules placed between different (mismatched) metals.
In both cases, the temperature profile generated on the current-carrying junction was spatially asymmetric. In junctions with asymmetric endgroups, a large temperature gradient developed on the weak contact, yet the diode effect when measured on the junction (rather than over the molecule) was marginal. 
For Au-alkane-Ag mismatched  junctions, the Au contact was more resistive, with a larger thermal bias falling on that contact. 
In the mismatched case, diode effects could be materialized, albeit at high biases.
We thus conclude that alkane-based junctions with mismatched metals show a thermal diode effect once large thermal biases (over 300 K) are applied. These conditions are currently beyond current experimental capabilities.

Moving beyond alkane-based chains, which serve here as a benchmark, our future work will be focused on the simulation of phonon heat transport in families of flexible molecules that are expected to show nonlinear transport behavior at moderate  thermal biases, thus possibly support a stronger diode response. 
Besides nonlinear effects, there is an interest in identifying families of molecules that either promote \cite{Rubtsov-ballistic} or hinder \cite{Gemma,Hatef21} thermal transport in single-molecule junctions. 

Other fundamental challenges in molecular thermal transport include:
(i) Performing a quantitative study of the relationship between transient pump-probe vibrational energy transfer experiments \cite{Troe04,Rubtsov19,Rubtsov21} and steady state measurements of phononic heat transport. (ii) Identifying means for an active control of thermal transport, e.g., with electric fields \cite{LeitnerE} or by mechanical compression \cite{GemmaM}, and  (iii) understanding whether and when quantum effects contribute to steady-state phonon transport at room temperature \cite{Leitner15,Paulyinter,Hatef19,Nitzan20inter,Hanna,Lu2021}. 

%
\begin{acknowledgments}
DS acknowledges the NSERC discovery grant and the Canada Research Chair Program. AJHM acknoledges National Science Foundation Award DMR-2025013.
\end{acknowledgments}


\renewcommand{\theequation}{A\arabic{equation}}
\renewcommand{\thesection}{A\arabic{section}}
\renewcommand{\thesubsection}{A\arabic{subsection}}
\setcounter{equation}{0}
\setcounter{section}{0} 
\setcounter{subsection}{0} 
\section*{Appendix A: Implementation of RNEMD simulations}\label{app:1}

We describe here technical details concerning the RNEMD method.
We present the setup and simulations for the 
junction. One can readily generalize the setups to account for a modified sulfur endgroup, S', and  use other metal contacts.

\subsection{Setup}
 \begin{figure}[hbt!]
 \centering
 \includegraphics[width=\columnwidth]{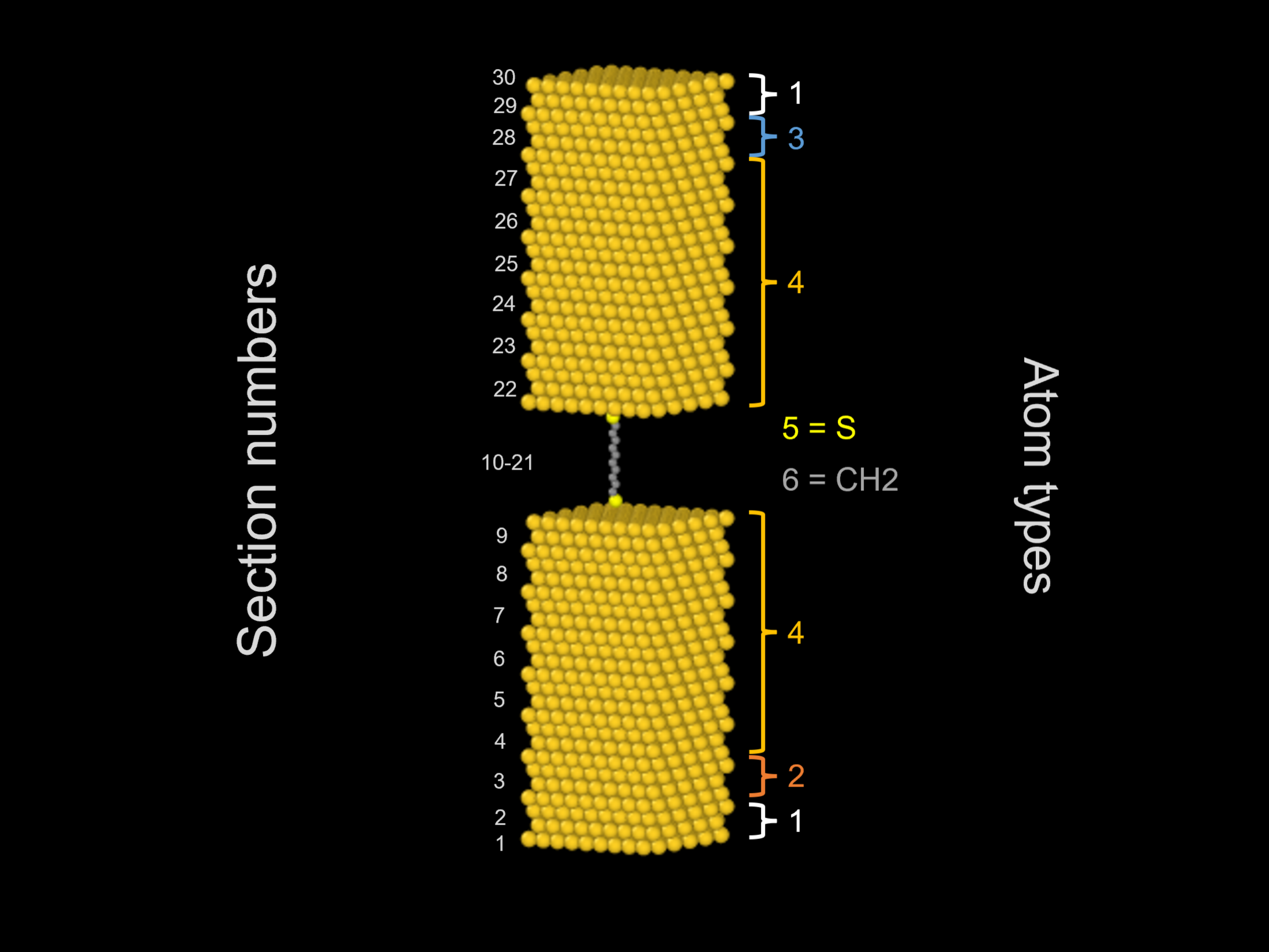}
 \caption{Visualization of the initial coordinate file (OVITO) used in RNEMD simulations for $N_{\text{C}}=10$. 
 We list atom types (1 to 6) and section numbers (1 to 30), with the temperature profile evaluated for each section (barring the boundaries, type 1).}
        \label{NEsetup}
 \end{figure}

The initial, rigid structure of the system is displayed in Fig.~\ref{NEsetup}, where a single alkanedithol chain is placed between two leads of gold with 2160 atoms each.  
The system is placed in a simulation box with periodic boundaries that has the $x$, $y$ dimensions conform to the edges of Au leads, but a substantial height in the $z$ dimension, with the top of the box placed far from the atoms.

The atom types comprise of Au, S, and CH$_2$. For this model, C and H are grouped into an united CH$_2$ atom due to negligible contribution to thermal transport by interactions from the C-H bond. Henceforth, C atoms will refer to united CH$_2$ atoms. 
We display in Fig.~\ref{NEsetup} our assignment of different atom types (1 to 6), as well as defined sections (1 to 30) for which temperature data will be collected. 
Note that we distinguish between four different Au atom (types 1 to 4) since they are handled differently in simulations, as we now explain.
Each Au section consists of 270 atoms (separated into sections with 180 and 90 atoms at the very ends). 
The individual S and C atoms on the molecule also define sections, giving a total of 30 sections for a 10-unit S-alkane-S chain.  
The total number of Au atoms in sections 3-to-9 is 1890, thus we have 3780 Au atoms in both leads. 
Our standard setup consists of 7 sections of Au at each contact, for which temperature data are collected for.

Shown in Fig.~\ref{NEsetup}, atoms type 1 exist on both ends of the leads as a fixed Au that holds the rest of the lead in place to prevent collapsing. 
These atoms are not allowed to move in simulations and their role is to hold the junction and prevent the gold pieces from collapsing onto one another.
Atom types 2 and 3 are single sections of Au, where heat is  inputted and outputted, respectively. The rest of the moving Au are grouped as type 4. 
Atom types 5 and 6 correspond to S and C, respectively, where we use the united atom description.

The force field and its parameters is described in Table~\ref{Table2}; it was adopted from Ref. \citenum{Ong14}.
Intermolecular potentials were adopted from simulations on nanocrystal arrays\cite{Ong14} and SAMs \cite{Shub15,Shub17}. Interactions within the alkanedithiol molecule are approximated by a harmonic two- and three-body interactions, as well as four-body (dihedral) interactions. The Morse potential governs the interaction between the metal and the molecule, namely the Au and S interaction. Au-Au interactions are given by the embedded atom model \cite{EAM1,EAM2}. A Lennard-Jones potential is set for long-range, non-bonded interaction between all atoms. 

\subsection{Equilibration}
NPT equilibration is carried out first to relax the system, barostatting for zero pressure and to a target temperature. 
NVT propagation is subsequently carried out by equilibrating the entire system to $T=300$ K. Each equilibration is carried over 1.5 ns with 1 fs timesteps. 

\subsection{Production Run}
After equilibration, we carry out long production runs with NVE simulations to produce temperature profiles in steady state. 
At this stage, heat is inputted and extracted (as kinetic energy per unit time to the atoms) at the same rate to and from atom types 2 and 3, respectively. It was determined that a heat current ($J$) on the order of 0.01-0.04 eV/ps produces temperature differences of order 50-200 K. Production runs are executed for a total of 22.5 ns, with results being averaged over the last 20 ns.
Raw temperature data is collected by logging every 1000 steps of 1 fs timesteps, equivalent to every 1 ps. 


\renewcommand{\theequation}{B\arabic{equation}}
\renewcommand{\thesection}{B\arabic{section}}
\renewcommand{\thesubsection}{B\arabic{subsection}}
\setcounter{equation}{0}
\setcounter{section}{0} 
\setcounter{subsection}{0} 
\section*{Appendix B: Implementation of AEMD simulations}
\label{app:2}
In this Appendix, we describe the implementation of the AEMD simulation method.

\begin{figure}[htbp]
\centering
\includegraphics[width=\columnwidth]{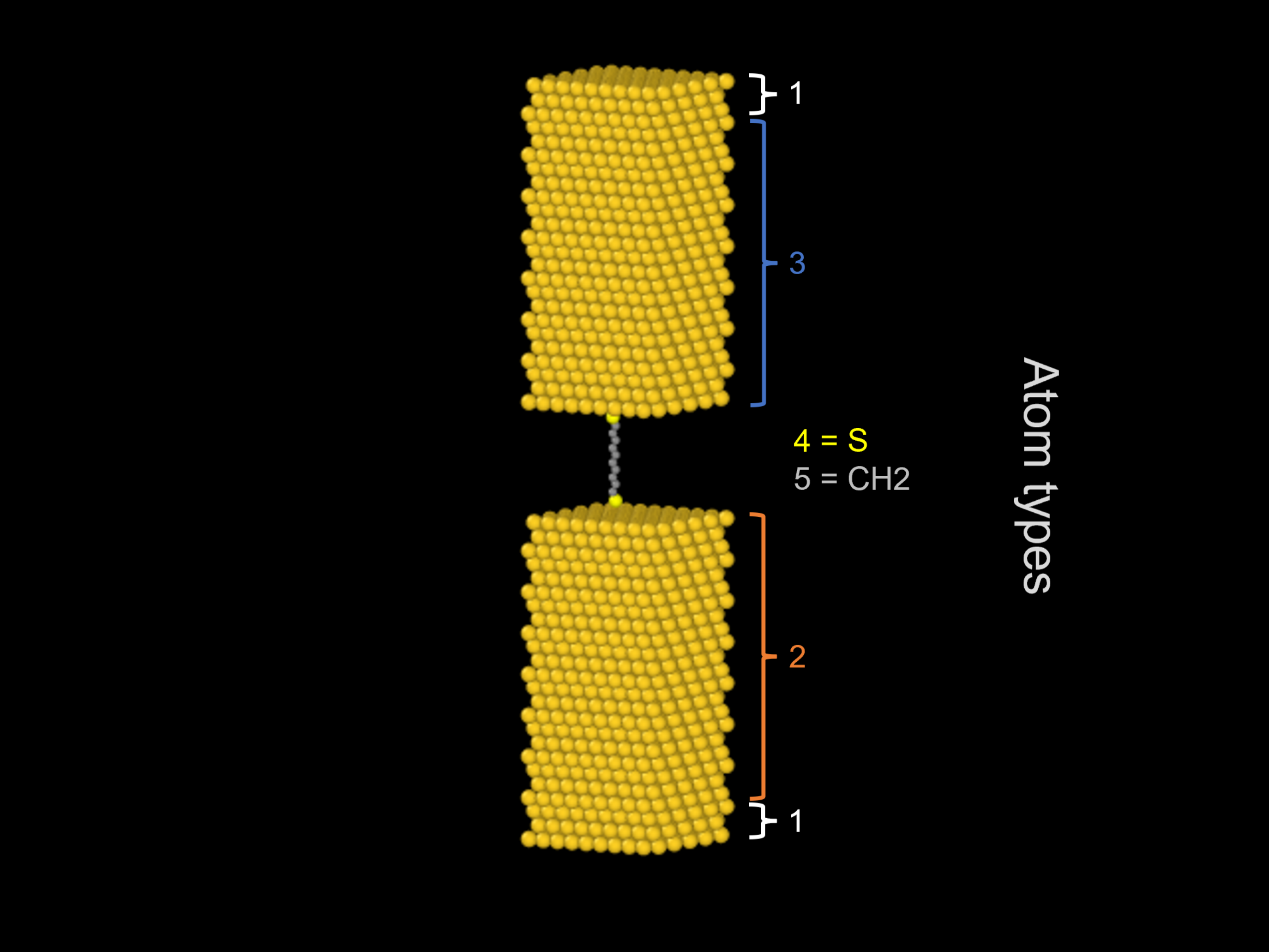} 
\caption{Visualization of AEMD initial coordinate file (OVITO) for $N_{\text{C}}=10$, with listed atom types used for simulation.}
\label{AEsetup}
\end{figure}
\subsection{Setup}

The initial, rigid structure of the system is displayed in Fig.~\ref{AEsetup}, where a single alkanedithol chain is placed between two leads of gold with 2160 atoms each. Similarly to RNEMD, the system has periodic boundaries in the $x$ and $y$ dimensions (planes of gold), but the box is extended in the $z$ dimension; the top of box is placed further away from the atoms.

Similarly to Appendix A, atom types comprise of Au, S, and CH$_2$. 
Shown in Fig.~\ref{AEsetup}, atom type 1 exists on both ends of the leads as the fixed Au that holds the rest of the lead in place to prevent collapsing.  
Atomss type 2 and 3 designate the moving Au of their respective leads, which will be equilibrated to high and low temperatures. 
Types 4 and 5 are S and C atoms, respectively.
The force field used was the same as in the RNEMD.

\subsection{Equilibration}

NPT equilibration is carried out first to relax the system and simulation box, barostatting for zero pressure and thermostatting to a target temperature, $\bar{T}$. 
Subsequent NVT equilibration is carried out individually on the two Au leads to $T_h$ and $T_c$, 
where $\bar{T}=\frac{1}{2}(T_h+T_c)$. Equilibration steps are carried out over 1.5 ns with 1 fs timesteps.

\subsection{Production Run}
The two leads are connected via the molecule.
An NVE simulation is conducted on the overall system for a total simulation time of 5 ns. 
During this time, the temperatures of the Au lead relax to the equilibrium value, $\bar{T}$. 
Raw temperature data is collected by logging every 1000 steps of 1 fs timesteps, equivalent to every 1 ps.

\begin{widetext}
\renewcommand{\theequation}{C\arabic{equation}}
\setcounter{equation}{0}
\setcounter{section}{0} 
\section*{Appendix C: Force field: Potentials and Parameters}\label{app:3}

See Table \ref{Table2}.

\begin{table*}[ht]
	\centering
	\caption[List of potentials and interactions] {List of potentials and interactions used in MD simulations.}
	\begin{tabular}{| m{20em} | m{7em} | m{15em} |}
		\hline
		\bf Potential & \bf Interaction & \bf Parameters\\
		\hline
		Harmonic - stretching  & C-C & $k_s=11.27$ eV/\AA$^2$, $r_0 = 1.54$ \AA\\
		$E_s=\cfrac{1}{2}k_s(r-r_0)^2$ & S-C & $k_s=9.67$ eV/\AA$^2$, $r_0 = 1.815$ \AA\\
		& & \\
		\hline
		Harmonic- bending & C-C-C & $k_\theta = 5.388$ eV/rad$^2$, $\theta_0 = 109.5^\circ$\\
		$E_\theta=\cfrac{1}{2}k_\theta(\theta-\theta_0)^2$ & S-C-C & $k_\theta = 5.388$ eV/rad$^2$, $\theta_0 = 114.4^\circ$\\
		& & \\
		\hline
		Dihedral  & C-C-C-C & $a_0 = 0.09617$ eV\\
		$E_d=\sum\limits_{n=0}^{5}a_n\cos^n\phi$ & S-C-C-C & $a_1 = -0.125988$ eV \\
		& (Same & $a_2 = -0.13598$ eV\\
		& parameters) & $a_3 = 0.0317$ eV\\
		& & $a_4 = 0.27196$ eV\\
		& & $a_5 = 0.32642$ eV\\
		& & \\
		\hline
		Morse & Au-S & $D_e = 0.38$ eV\\
		$E_M=D_e[e^{-2\alpha(r-r_0)}-2e^{-\alpha (r-r_0)}]$ & &$\alpha = 1.47$ \AA$^{-1}$\\
		& & $r_0 = 2.65$ \AA\\
		& & \\
		\hline
		Embedded atom model & Au-Au & Ref.\cite{EAM1} \\
		& Ag-Ag, Au-Ag & Ref.\cite{EAM2}\\
		& &\\
		\hline
		Lennard-Jones &  Au and Au & $\varepsilon = 0.00169$ eV, $\sigma =2.935$ \AA\\
		$E_{LJ}=4\varepsilon_{ij}\left[\left(\frac{\sigma_{ij}}{r_{ij}}\right)^{12}-\left(\frac{\sigma_{ij}}{r_{ij}}\right)^6\right]$ & Au and C & $\varepsilon = 0.00294$ eV, $\sigma =3.42$ \AA\\
		with Lorentz-Berthelot mixing rules & S and S & $\varepsilon = 0.01724$ eV, $\sigma =4.25$ \AA\\
		$\varepsilon_{ij}=\sqrt{\varepsilon_i\varepsilon_j}, \quad \sigma_{ij}=\frac{1}{2}(\sigma_i+\sigma_j)$ & S and C & $\varepsilon = 0.01086$ eV, $\sigma =3.55$ \AA\\
		where $i,j$ denote different atom types & C and C & $\varepsilon = 0.00512$ eV, $\sigma =3.905$ \AA\\
		& &\\
		\hline
	\end{tabular}
	\label{Table2}
\end{table*}

\vspace{4mm}

\begin{figure*}[htb]
\centering
\includegraphics[width=\columnwidth]{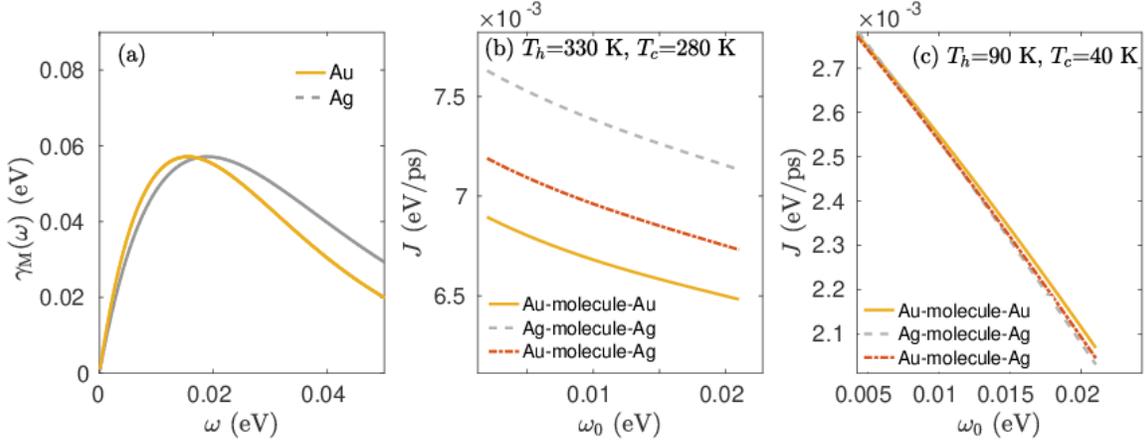} 
\caption{Quantum analysis of phonon heat transport in M-alkanedithiol-M' junctions using a single-mode model.
{\bf (a)} Phononic spectral density functions $\gamma_M(\omega)$ used to describe M=Au,Ag, differing by their Debye frequencies,
$T_D^{\text{ Au}}=180$ K and $T_D^{\text{ Ag}}=220$ K, and normalized by height.
{\bf (b)} Heat current for gold, silver, and mismatched junctions for $T_h=340$ K, $T_c=270$ K and
at {\bf (c)} lower temperatures, $T_h=90$ K, $T_c=40$ K.}
\label{Debye1}
\end{figure*}

\begin{figure*}[htb]
\centering
\includegraphics[width=0.7\columnwidth]{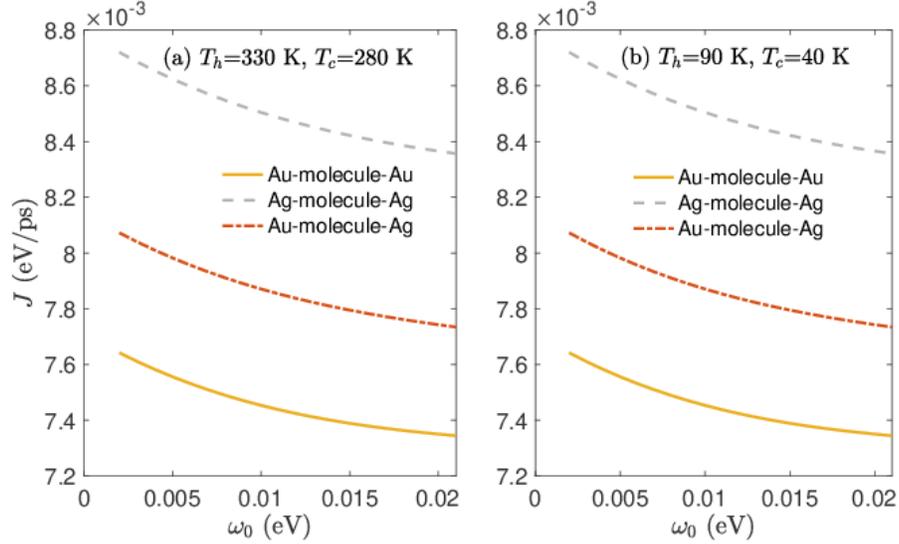} 
\caption{Analysis of phonon heat transport in M-alkanedithiol-M' junctions based on a single-mode model, as in Fig. \ref{Debye1}, but using here the classical high-temperature limit for the baths occupation functions.
 Heat current for gold, silver, and mismatched junctions at {\bf (a)} $T_h=340$ K, $T_c=270$ K and
 {\bf (b)} lower temperatures, $T_h=90$ K, $T_c=40$ K.}
\label{Debye1C}
\end{figure*}

\renewcommand{\theequation}{D\arabic{equation}}
\setcounter{equation}{0}
\setcounter{section}{0} 
\setcounter{subsection}{0} 
\section*{Appendix D: Heat transport in M-molecule-M' junctions: Landauer formalism}
\label{app:4}


We support here the MD results presented in Fig. \ref{AuAg_bar} using a minimal model for thermal transport.
Our goal is to demonstrate the effect of Debye frequencies of the attached metals (or more generally, their phonon spectra) on the phonon heat current.
Using the quantum Landauer formula for phonon heat transport, we study here heat transport in M-alkanedithiol-M' junctions with the metals being either Au or Ag. Consistent with Fig. \ref{AuAg_bar}, we find that around room temperature, Ag, which has a higher Debye frequency than Au, supports higher currents for the same thermal bias. Mismatched junctions show conductance values in between the same-metal junctions.

In our minimal model, the molecule is represented by a single harmonic mode of frequency $\omega_0$. We assume Ohmic functions for the two phonon baths,  albeit we use different Debye temperatures for Au and Ag. To restrict the comparison on the difference in Debye frequencies, we normalize the heights of the two spectral functions so as they match at their maximum value.

For a fully harmonic model, the phonon heat current is given by a Landauer-type expression \cite{Dhar}. Considering a   single-mode system, a calculation of the transmission function yields \cite{Dvira2003, SB1}
%
\bea
J=\frac{2}{\pi} \int_{-\infty}^{\infty}
 d\omega \hbar \omega\frac{ \omega^2 \gamma_{\text M}(\omega)\gamma_{\text{M'}}(\omega)}{ (\omega^2-\omega_0^2)^2 + \omega^2(\gamma_{\text M}(\omega)+\gamma_{\text{M'}}(\omega))^2}
\left[ n_{\text M}(\omega)-n_{\text {M'}}(\omega)\right].
\nonumber\\
\eea
%
Here, $n_{\text M}(\omega)$ is the Bose-Einstein distribution function evaluated at the temperature of the M metal. $\gamma_M(\omega)=C_M\omega e^{-\omega/\omega_M}$
with $C_M$ a dimensionless constant controlling the metal-molecule coupling strengths. $\omega_M$ is the Debye frequency of the M metal. The two spectral functions are presented in Fig. \ref{Debye1}(a). Since $\omega_{\text Ag}>\omega_{\text Au}$,  the phonon spectra of Ag is displaced to higher frequencies.

Our results for the heat current are presented in Fig. \ref{Debye1}(b)-(c). Since we take into account a single molecular mode, we study the current as a function of the
molecular frequency $\omega_0$. The behavior of the current  is monotonic with $\omega_0$, thus the overall trends are expected to hold even when multiple modes contribute. 

Around room temperature, according to Fig. \ref{Debye1}(b),
Ag junctions  support the highest currents, Au junctions have the lowest currents, and the mismatched case shows an in-between behavior. These results qualitatively agree with and support RNEMD simulations as presented in Fig. \ref{AuAg_bar}. 
This behavior can be rationalized based on the higher spectral density of Ag metal compared to Ag around room temperature, 0.025 eV. Only when we substantially reduce the temperature, thus probing the low-frequency regime of the spectral functions, the reversed trend is observed, with gold junctions delivering the highest conductance, see Fig. \ref{Debye1}(c).

To decouple in our model harmonic effects from quantum statistics, 
in Figure \ref{Debye1C} we replace the quantum baths by classical ones. That is, we replace the Bose-Einstein distribution functions by their classical-high temperature limits. From the comparison between Fig.
\ref{Debye1} and \ref{Debye1C} we learn that 
while transport around room temperature is only mildly affected by quantum effects (in this fully harmonic analysis), at lower temperatures quantum effects become significant, and the classical current overestimates the current. We can also safely conclude that the
ordering $G_{\rm Ag-Ag}> G_{\rm Ag-Au}> G_{\rm Au-Au}$ can be rationalized based on a classical harmonic model and emerging from the Debye temperature of silver being higher than gold's.

As for a quantitative comparison between the minimal-model calculation and classical MD simulations: Given the significant simplifications in the minimal model here (single molecular mode, fully harmonic force field) we do not expect the numbers to agree. 
However, we note that according to the MD study the conductance of Ag junctions is about 40\% higher than Au, while mismatched junctions lie in between, with 10\% enhancement of conductance relative to Au, see Fig. \ref{AuAg_bar}. The single-model model provide smaller ratios, with $G_{\text {Ag-Ag}}/G_{\text{Au-Au}}\approx 1.1$, see  Fig. \ref{Debye1}(b).

\end{widetext}




\begin{thebibliography}{99}




\bibitem{Pop10}
E. Pop,
"Energy dissipation and transport in nanoscale devices,"
\href{https://doi.org/10.1007/s12274-010-1019-z}{Nano Res.} {\bf 3}, 147 (2010).

\bibitem{Baowen12}
N. Li, J. Ren, L. Wang, G. Zhang, P. H\"anggi, and B. Li,
"Colloquium: Phononics: Manipulating heat flow with electronic analogs and beyond,"
\href{https://doi.org/10.1103/RevModPhys.84.1045}{Rev. Mod. Phys.} {\bf 84}, 1045 (2012).

\bibitem{Luo13}
T. Luo and G. Chen, 
"Nanoscale Heat Transfer - from Computation to Experiment," 
\href{https://doi.org/10.1039/C2CP43771F}{Phys. Chem. Chem. Phys.} {\bf 15}, 3389 (2013).

\bibitem{Rev14}
D. G. Cahill,  P. V. Braun, G. Chen, D. R. Clarke, S. Fan, K. E. Goodson, P. Keblinski, W. P. King, G. D. Mahan, A. Majumdar, H. J. Maris, S. R. Phillpot, E. Pop, and L. Shi,
"Nanoscale thermal transport. II. 2003–2012,"
\href{https://doi.org/10.1063/1.4832615}{Appl. Phys. Rev.} {\bf 1}, 011305 (2014).

\bibitem{Leitner15}
D. M. Leitner, 
"Quantum Ergodicity and Energy Flow in Molecules,"
\href{https://doi.org/10.1080/00018732.2015.1109817}{Adv. Phys.} {\bf 64}, 445 (2015).

\bibitem{RevA}
D. Segal and  B. K. Agarwalla,
"Vibrational Heat Transport in Molecular Junctions,"
\href{https://doi.org/10.1146/annurev-physchem-040215-112103}{Ann. Rev. Phys. Chem.} {\bf 67}, 185 (2016).

\bibitem{Yoon20}
S. Park, J. Jang, H. Kim, D. I. Park, K. Kim, and H. J. Yoon,
"Thermal conductance in single molecules and self-assembled monolayers: physicochemical insights, progress, and challenges,"
\href{https://doi.org/10.1039/D0TA07095E}{J. Mater. Chem. A} {\bf 8}, 19746 (2020).

\bibitem{BaowenR21}
Y. Li, W. Li, T. Han, X. Zheng, J. Li, B. Li, S. Fan, and C.-W. Qiu,
"Transforming heat transfer with thermal metamaterials and devices,"
\href{https://doi.org/10.1038/s41578-021-00283-2}{Nat. Rev. Mater.} {\bf 6}, 488 (2021).

\bibitem{RevG}
B. Gotsmann, A. Gemma, and D. Segal,
"Quantum phonon transport through channels and molecules—A Perspective,"
\href{https://doi.org/10.1063/5.0088460}{App. Phys. Lett.} {\bf 120}, 160503 (2022).


\bibitem{Dvira2003}
D. Segal, A. Nitzan, and P. H\"anggi,
"Thermal conductance through molecular wires,"
\href{https://doi.org/10.1063/1.1603211}{J. Chem. Phys.} {\bf 119}, 6840 (2003).

\bibitem{Pawel11}
K. Sasikumar and P. Keblinski,
"Effect of chain conformation in the phonon
transport across a Si-polyethylene single-
molecule covalent junction,"
\href{https://doi.org/10.1063/1.3592296}{J. Appl. Phys.} {\bf 109}, 114307 (2011). 

\bibitem{Pauly16}
J. C. Kl\"ockner, M. B\"urkle, J. C. Cuevas, and F. Pauly,
"Length dependence of the thermal conductance of alkane-based single-molecule junctions: An ab initio study,"
\href{https://doi.org/10.1103/PhysRevB.94.205425}{Phys. Rev. B} {\bf 94}, 205425 (2016).

\bibitem{Pauly18}
J. C. Kl\"ockner, J. C. Cuevas, and F. Pauly,
"Transmission eigenchannels for coherent phonon transport,"
\href{https://doi.org/10.1103/PhysRevB.97.155432}{Phys. Rev. B} {\bf 97}, 155432 (2018).

\bibitem{Roya19}
R. Moghaddasi Fereidani and D. Segal,
"Phononic heat transport in molecular junctions: Quantum effects and vibrational mismatch,"
\href{https://doi.org/10.1063/1.5075620}{J. Chem. Phys.} {\bf 150}, 024105 (2019).

\bibitem{Nitzan20}
I. Sharony, R. Chen, and A. Nitzan,
"Stochastic simulation of nonequilibrium heat conduction in extended molecular junctions,"
\href{https://doi.org/10.1063/5.0022423}{J. Chem. Phys.} {\bf 153}, 144113 (2020).

\bibitem{Lu2021}
G. Li, B.-Z. Hu, N. Yang, and J.-T. L\"u,
"Temperature-dependent thermal transport of single molecular junctions from semiclassical Langevin molecular dynamics,"
\href{https://doi.org/10.1103/PhysRevB.104.245413}{Phys. Rev. B} {\bf 104}, 245413 (2021).

\bibitem{Nitzan22}
M. Dinpajooh and A. Nitzan,
"Heat conduction in polymer chains: Effect of substrate on the thermal conductance,"
\href{https://doi.org/10.1063/5.0087163}{J. Chem. Phys.} {\bf 156}, 144901 (2022).

\bibitem{CuiExp19}
L. Cui, S. Hur, Z. A. Akbar, J. C. Kl\"ockner, W. Jeong, F. Pauly, S.-Y. Jang, P. Reddy, and E. Meyhofer,
"Thermal conductance of single-molecule junctions,"
\href{https://doi.org/10.1038/s41586-019-1420-z}{Nature} {\bf 572}, 628 (2019).

\bibitem{GotsmannExp19}
N. Mosso, H. Sadeghi, A. Gemma, S. Sangtarash, U. Drechsler, C. Lambert, and B. Gotsmann,
"Thermal Transport through Single-Molecule Junctions,"
\href{https://doi.org/10.1021/acs.nanolett.9b02089}{Nano Lett.} {\bf 19}, 7614 (2019).


\bibitem{Wang06}
R. Y. Wang, R. A. Segalman, and A. Majumdar,
"Room temperature thermal conductance of alkanedithiol self-assembled monolayers,"
\href{https://doi.org/10.1063/1.2358856}{Appl. Phys. Lett.} {\bf 89}, 173113 (2006).

\bibitem{Dlott07}
Z. Wang, J. A. Carter, A. Lagutchev, Y. K. Koh, N.-H. Seong, D. G. Cahill, and D. D. Dlott,
"Ultrafast Flash Thermal Conductance of Molecular Chains,"
\href{https://doi.org/10.1126/science.1145220}{Science} {\bf 317}, 787 (2007).

\bibitem{Cahill12}
M. D. Losego, M. E. Grady, N. R. Sottos, D. G. Cahill, and P. V. Braun,
"Effects of chemical bonding on heat transport across interfaces,"
\href{https://doi.org/10.1038/nmat3303}{Nat. Mater.} {\bf 11}, 502 (2012).

\bibitem{GotsmannExp14}
T. Meier, F. Menges, P. Nirmalraj, H. H\"olscher, H. Riel, and B. Gotsmann,
"Length-Dependent Thermal Transport along Molecular Chains,"
\href{https://doi.org/10.1103/PhysRevLett.113.060801}{Phys. Rev. Lett.} {\bf 113}, 060801 (2014).

\bibitem{Shub15}
S. Majumdar, J. A. Sierra-Suarez, S. N. Schiffres, W.-L. Ong, C. F. Higgs III, A. J. H. McGaughey, and J. A. Malen,
"Vibrational Mismatch of Metal Leads Controls Thermal Conductance of Self-Assembled Monolayer Junctions,"
\href{https://doi.org/10.1021/nl504844d}{Nano Lett.} {\bf 15}, 2985 (2015).

\bibitem{Shub17}
S. Majumdar, J. A. Malen, and A. J. H. McGaughey,
"Cooperative Molecular Behavior Enhances the Thermal Conductance of Binary Self-Assembled Monolayer Junctions,"
\href{https://doi.org/10.1021/acs.nanolett.6b03894}{Nano Lett.} {\bf 17}, 220 (2017).

\bibitem{Luo10}
T. Luo and J. R. Lloyd,
"Non-equilibrium molecular dynamics study of thermal energy transport in Au–SAM–Au junctions,"
\href{https://doi.org/10.1016/j.ijheatmasstransfer.2009.10.033}{J. Heat Mass Transf.} {\bf 53}, 1 (2010).

\bibitem{Hu10}
L. Hu, L. Zhang, M. Hu, J.-S. Wang, B. Li, and P. Keblinski,
"Phonon interference at self-assembled monolayer interfaces: Molecular dynamics simulations,"
\href{https://doi.org/10.1103/PhysRevB.81.235427}{Phys. Rev. B} {\bf 81}, 235427 (2010).

\bibitem{Kikugawa14}
G. Kikugawa, T. Ohara, T. Kawaguchi, I. Kinefuchi, and Y. Matsumoto,
"A molecular dynamics study on heat conduction characteristics inside the alkanethiolate SAM and alkane liquid,"
\href{https://doi.org/10.1016/j.ijheatmasstransfer.2014.07.040}{J. Heat Mass Transf.} {\bf 78}, 630 (2014).

\bibitem{Diamond17}
Y. Wang, Y. Cao, K. Zhou, and Z. Xu,
"Assessment of Self-Assembled Monolayers as High-Performance Thermal Interface Materials,"
\href{https://doi.org/10.1002/admi.201700355}{Adv. Mater. Interfaces} {\bf 4}, 1700355 (2017).



\bibitem{Quantum1}
J.-S. Wang, B. K. Agarwalla, H. Li, and J. Thingna, 
"Nonequilibrium Green’s function method for quantum thermal transport,"
\href{https://doi.org/10.1007/s11467-013-0340-x}{Front. Phys.} {\bf 9}, 673 (2014).

\bibitem{Quantum2}
Y.-J. Zeng, Z.-K. Ding, H. Pan, Y.-X. Feng and K.-Q. Chen,
"Nonequilibrium Green's function method for phonon heat transport in quantum system,"
\href{https://doi.org/10.1088/1361-648X/ac5c21}{J. Phys.: Condens. Matter} {\bf 34}, 223001 (2022).

\bibitem{LAMMPS}
A. P. Thompson, H. M. Aktulga, R. Berger, D. S. Bolintineanu, W. M. Brown, P. S. Crozier, P. J. in 't Veld, A. Kohlmeyer, S. G. Moore, T. D. Nguyen, R. Shan, M. J. Stevens, J. Tranchida, C. Trott, and S. J. Plimpton,
"LAMMPS - a flexible simulation tool for particle-based materials modeling at the atomic, meso, and continuum scales,"
\href{https://doi.org/10.1016/j.cpc.2021.108171}{Comp. Phys. Comm.} {\bf 271}, 10817 (2022).

\bibitem{AEMD}
T.-Q. Duong, C. Massobrio, G. Ori, M. Boero, and E. Martin,
"Thermal resistance of an interfacial molecular layer by first-principles molecular dynamics,"
\href{https://doi.org/10.1063/5.0014232}{J. Chem. Phys.} {\bf 153}, 074704 (2020).


\bibitem{WalkerRev11}
N. A. Roberts and D. G. Walker,
"A review of thermal rectification observations and models in solid materials,"
\href{https://doi.org/10.1016/j.ijthermalsci.2010.12.004}{Int. J. Therm. Sci.} {\bf 50}, 648 (2011).

\bibitem{DiodeRev17}
G. Wehmeyer, T. Yabuki, C. Monachon, J. Wu, and C. Dames,
"Thermal diodes, regulators, and switches: Physical mechanisms and potential applications,"
\href{https://doi.org/10.1063/1.5001072}{Appl. Phys. Rev.} {\bf 4}, 041304 (2017).

\bibitem{Terraneo02}
M. Terraneo, M. Peyrard, and G. Casati,
"Controlling the Energy Flow in Nonlinear Lattices: A Model for a Thermal Rectifier,"
\href{https://doi.org/10.1103/PhysRevLett.88.094302}{Phys. Rev. Lett.} {\bf 88}, 094302 (2002).

\bibitem{Baowen04}
B. Li, L. Wang, and G. Casati,
"Thermal Diode: Rectification of Heat Flux,"
\href{https://doi.org/10.1103/PhysRevLett.93.184301}{Phys. Rev. Lett.} {\bf 93}, 184301 (2004).

\bibitem{Bambi06}
B. Hu, L. Yang, and Y. Zhang,
"Asymmetric Heat Conduction in Nonlinear Lattices,"
\href{https://doi.org/10.1103/PhysRevLett.97.124302}{Phys. Rev. Lett.} {\bf 97}, 124302 (2006).

\bibitem{SB1}
D. Segal and A. Nitzan,
"Spin-Boson Thermal Rectifier,"
\href{https://doi.org/10.1103/PhysRevLett.94.034301}{Phys. Rev. Lett.} {\bf 94}, 034301 (2005).

\bibitem{SB2}
L.-A. Wu, C. X. Yu, and D. Segal,
"Nonlinear quantum heat transfer in hybrid structures: Sufficient conditions for thermal rectification,"
\href{https://doi.org/10.1103/PhysRevE.80.041103}{Phys. Rev. E} {\bf 80}, 041103 (2009).

\bibitem{NEMD19}
Z. Li, S. Xiong, C. Sievers, Y. Hu, Z. Fan, N. Wei, H. Bao, S. Chen, D. Donadio, and T. Ala-Nissila,
"Influence of thermostatting on nonequilibrium molecular dynamics simulations of heat conduction in solids,"
\href{https://doi.org/10.1063/1.5132543}{J. Chem. Phys.} {\bf 151}, 234105 (2019).

\bibitem{Lebo67}
Z. Rieder, J. L. Lebowitz, and E. Lieb,
"Properties of a Harmonic Crystal in a Stationary Nonequilibrium State,"
\href{https://doi.org/10.1063/1.1705319}{J. Math. Phys.} {\bf 8}, 1073 (1967). 

\bibitem{BaowenR}
J. Chen, X. Xu, J. Zhou, and B. Li,
"Interfacial thermal resistance: Past, present, and future,"
\href{https://doi.org/10.1103/RevModPhys.94.025002}{Rev. Mod. Phys.} {\bf 94}, 025002 (2022).

\bibitem{BLD2}
G. Wu and B. Li,
"Thermal rectification in carbon nanotube intramolecular junctions: Molecular dynamics calculations,"
\href{https://doi.org/10.1103/PhysRevB.76.085424}{Phys. Rev. B} {\bf 76}, 085424 (2007).

\bibitem{BLD3}
G. Wu and B. Li,
"Thermal rectifiers from deformed carbon nanohorns,"
\href{https://doi.org/10.1088/0953-8984/20/17/175211}{J. Phys.: Condens. Matter} {\bf 20}, 175211 (2008).

\bibitem{BLD4}
M. Hu, P. Keblinski, and B. Li,
"Thermal rectification at silicon-amorphous polyethylene interface,"
\href{https://doi.org/10.1063/1.2937834}{Appl. Phys. Lett.} {\bf 92}, 211908 (2008).

\bibitem{BLD5}
N. Yang, G. Zhang, and B. Li,
"Carbon nanocone: A promising thermal rectifier,"
\href{https://doi.org/10.1063/1.3049603}{Appl. Phys. Lett.} {\bf 93}, 243111 (2008).

\bibitem{BLD6}
N. Yang, G. Zhang, and B. Li,
"Thermal rectification in asymmetric graphene ribbons,"
\href{https://doi.org/10.1063/1.3183587}{Appl. Phys. Lett.} {\bf 95}, 033107 (2009).

\bibitem{BLD7}
L. Zhang, J.-T. L\"u, J.-S. Wang, and B. Li,
"Thermal transport across metal-insulator interface via electron-phonon interaction,"
\href{https://doi.org/10.1088/0953-8984/25/44/445801}{J. Phys.: Condens. Matter} {\bf 25}, 445801 (2013).

\bibitem{rad1}
D. Segal,
"Single Mode Heat Rectifier: Controlling Energy Flow Between Electronic Conductors,"
\href{https://doi.org/10.1103/PhysRevLett.100.105901}{Phys. Rev. Lett.} {\bf 100}, 105901 (2008).
%

\bibitem{rad2}
F. Giazotto and F. S. Bergeret,
"Thermal rectification of electrons in hybrid normal metal-superconductor nanojunctions,"
\href{https://doi.org/10.1063/1.4846375}{Appl. Phys. Lett.} {\bf 103}, 242602 (2013).

\bibitem{rad3}
E. Nefzaoui, K. Joulain, J. Drevillon, and Y. Ezzahri,
"Radiative thermal rectification using superconducting materials,"
\href{https://doi.org/10.1063/1.4868251}{Appl. Phys. Lett.} {\bf 104}, 103905 (2014).

\bibitem{Biehs13}
P. Ben-Abdallah and S.-A. Biehs,
"Phase-change radiative thermal diode,"
\href{https://doi.org/10.1063/1.4829618}{Appl. Phys. Lett.} {\bf 103}, 191907 (2013).

\bibitem{Xu15}
E. Pallecchi, Z. Chen, G. E. Fernandes, Y. Wan, J. H. Kim, and J. Xu,
"A thermal diode and novel implementation in a phase-change material,"
\href{https://doi.org/10.1039/C4MH00193A}{Mater. Horiz.} {\bf 2}, 125 (2015).

\bibitem{Fiorino18}
A. Fiorino, D. Thompson, L. Zhu, R. Mittapally, S.-A. Biehs, O. Bezencenet, N. El-Bondry, S. Bansropun, P. Ben-Abdallah, E. Meyhofer, and P. Reddy,
"A Thermal Diode Based on Nanoscale Thermal Radiation,"
\href{https://doi.org/10.1021/acsnano.8b01645}{ACS Nano} {\bf 12}, 5774 (2018).


\bibitem{Dlott13a}
B. C. Pein, Y. Sun, and D. D. Dlott,
"Unidirectional Vibrational Energy Flow in Nitrobenzene,"
\href{https://doi.org/10.1021/jp3127863}{J. Phys. Chem. A} {\bf 117}, 6066 (2013).

\bibitem{Dlott13b}
B. C. Pein, Y. Sun, and D. D. Dlott,  
"Controlling Vibrational Energy Flow in Liquid Alkylbenzenes," 
\href{https://doi.org/10.1021/jp406528u}{J. Phys. Chem. B} {\bf 117}, 10898 (2013). 

\bibitem{Tucker19}
A. J. Schmitz, H. D. Pandey, F. Chalyavi, T. Shi, E. E. Fenlon, S. H. Brewer, D. M. Leitner, and M. J. Tucker,
"Tuning Molecular Vibrational Energy Flow within an Aromatic Scaffold via Anharmonic Coupling,"
\href{https://doi.org/10.1021/acs.jpca.9b08010}{J. Phys. Chem. A} {\bf 123}, 10571 (2019).


\bibitem{LeitnerR}
K. M. Reid, H. D. Pandey, and D. M. Leitner,
"Elastic and Inelastic Contributions to Thermal Transport between Chemical Groups and Thermal Rectification in Molecules,"
\href{https://doi.org/10.1021/acs.jpcc.8b11640}{J. Phys. Chem. C} {\bf 123}, 6256 (2019).

\bibitem{Segal09}
D. Segal,
"Absence of thermal rectification in asymmetric harmonic chains with self-consistent reservoirs,"
\href{https://doi.org/10.1103/PhysRevE.79.012103}{Phys. Rev. E} {\bf 79}, 012103 (2009).

\bibitem{Rubtsov-ballistic}
N. I. Rubtsova, L. N. Qasim, A. A. Kurnosov, A. L. Burin, and I. V. Rubtsov,
"Ballistic Energy Transport in Oligomers,"
\href{https://doi.org/10.1021/acs.accounts.5b00299}{Acc. Chem. Res.} {\bf 48}, 2547 (2015).

\bibitem{Gemma}
Q. Li, M. Strange, I. Duchemin, D. Donadio, and G. C. Solomon, 
"A Strategy to Suppress Phonon Transport in Molecular Junctions Using $\pi$-Stacked Systems," 
\href{https://doi.org/10.1021/acs.jpcc.7b02005}{J. Phys. Chem. C} {\bf 121}, 7175 (2017).

\bibitem{Hatef21}
M. D. Noori, S. Sangtarash, and H. Sadeghi,
"The Effect of Anchor Group on the Phonon Thermal Conductance of Single Molecule Junctions,"
\href{https://doi.org/10.3390/app11031066}{Appl. Sci.} {\bf 11}, 1066 (2021).


\bibitem{Troe04}
D. Schwarzer, P. Kutne, C. Schr\"oder, and J. Troe,
"Intramolecular vibrational energy redistribution in bridged azulene-anthracene compounds: Ballistic energy transport through molecular chains," 
\href{https://doi.org/10.1063/1.1765092}{J. Chem. Phys.} {\bf 121}, 1754 (2004).

\bibitem{Rubtsov19}
I. V. Rubtsov and A. L. Burin, 
"Ballistic and diffusive vibrational energy transport in molecules," 
\href{https://doi.org/10.1063/1.5055670}{J. Chem. Phys.} {\bf 150}, 020901 (2019). 

\bibitem{Rubtsov21}
T. X. Leong, L. N. Qasim, R. T. Mackin, Y. Du, R. A. Pascal Jr., and  I. V. Rubtsov,
"Unidirectional coherent energy transport via conjugated oligo({\it p}-phenylene) chains," 
\href{https://doi.org/10.1063/5.0046932}{J. Chem. Phys.} {\bf 154}, 134304 (2021).


\bibitem{LeitnerE}
A. Maitra, S. Sarkar, D. M. Leitner, and J. M. Dawlaty, 
"Electric Fields Influence Intramolecular Vibrational Energy Relaxation and Line Widths,"
\href{https://doi.org/10.1021/acs.jpclett.1c02238}{J. Phys. Chem. Lett.} {\bf 12}, 7818 (2021).

\bibitem{GemmaM}
Q. Li, I. Duchemin, S. Xiong, G. C. Solomon, and D. Donadio,
"Mechanical Tuning of Thermal Transport in a Molecular Junction,"
\href{https://doi.org/10.1021/acs.jpcc.5b07429}{J. Phys. Chem. C} {\bf 119}, 24636 (2015).

\bibitem{Paulyinter}
J. C. Klöckner, J. C. Cuevas, and F. Pauly,
"Tuning the thermal conductance of molecular junctions with interference effects,"
\href{https://doi.org/10.1103/PhysRevB.96.245419}{Phys. Rev. B} {\bf 96}, 245419 (2017).

\bibitem{Hatef19}
H. Sadeghi,
"Quantum and Phonon Interference-Enhanced Molecular-Scale Thermoelectricity,"
\href{https://doi.org/10.1021/acs.jpcc.8b12538}{J. Phys. Chem. C} {\bf 123}, 12556 (2019).

\bibitem{Nitzan20inter}
R. Chen, I. Sharony, and A. Nitzan,
"Local Atomic Heat Currents and Classical Interference in Single-Molecule Heat Conduction,"
\href{https://doi.org/10.1021/acs.jpclett.0c00471}{J. Phys. Chem. Lett.} {\bf 11}, 4261 (2020).

\bibitem{Hanna}
P. Carpio-Martínez and G. Hanna, 
"Quantum bath effects on nonequilibrium heat transport in model molecular junctions," 
\href{https://doi.org/10.1063/5.0040752}{J. Chem. Phys.} {\bf 154}, 094108 (2021).

%
\bibitem{Ong14}
W.-L. Ong, S. Majumdar, J. A. Malen, and A. J. H. McGaughey,
"Coupling of Organic and Inorganic Vibrational States and Their Thermal Transport in Nanocrystal Arrays,"
\href{https://doi.org/10.1021/jp4120157}{J. Phys. Chem. C} {\bf 118}, 7288 (2014).

\bibitem{EAM1}
G. Grochola, S. P. Russo, and I. K. Snook,
"On fitting a gold embedded atom method potential using the force matching method,"
\href{https://doi.org/10.1063/1.2124667}{J. Chem. Phys.} {\bf 123}, 204719 (2005).

\bibitem{EAM2}
X. W. Zhou, R. A. Johnson, and H. N. G. Wadley,
"Misfit-energy-increasing dislocations in vapor-deposited CoFe/NiFe multilayers,"
\href{https://doi.org/10.1103/PhysRevB.69.144113}{Phys. Rev. B} {\bf 69}, 144113 (2004).


\bibitem{Dhar}
A. Dhar,
"Heat transport in low-dimensional systems,"
\href{https://doi.org/10.1080/00018730802538522}{Adv. Phys.} {\bf 57}, 457 (2008).

\end{thebibliography}
\end{document}